\documentstyle[emulateapj]{article}

\lefthead{} 
\righthead{OPTICAL SPECTROSCOPY OF XTE J1118+480}

\def\gsim{\;\rlap{\lower 2.5pt
 \hbox{$\sim$}}\raise 1.5pt\hbox{$>$}\;}
\def\lsim{\;\rlap{\lower 2.5pt
   \hbox{$\sim$}}\raise 1.5pt\hbox{$<$}\;} 

\begin{document}

\title{Optical Spectroscopy Of The X-Ray Transient XTE
J1118+480 In Outburst\altaffilmark{1}}

\author{Guillaume Dubus\altaffilmark{2}, Rita
S.J. Kim\altaffilmark{3}, Kristen Menou\altaffilmark{3,4}, Paula
Szkody\altaffilmark{5}\\ and David V. Bowen\altaffilmark{3}}

\altaffiltext{1}{Based on observations obtained with the
Apache Point Observatory 3.5-meter telescope, which is owned and
operated by the Astrophysical Research Consortium.}

\altaffiltext{2}{Astronomical Institute ``Anton Pannekoek'',
University of Amsterdam, Kruislaan 403, 1098 SJ, Amsterdam, the
Netherlands; gd@astro.uva.nl}

\altaffiltext{3}{Princeton University, Department of Astrophysical
Sciences, Princeton NJ 08544, USA; rita, kristen,
dvb@astro.princeton.edu}

\altaffiltext{4}{Chandra Fellow}

\altaffiltext{5}{Department of Astronomy, University of Washington,
Seattle, WA 98195; szkody@astro.washington.edu}

\begin{abstract}
We report on optical spectroscopic observations of the X-ray Transient
XTE~J1118+480 covering the period from April 7, 2000 to July 4,
2000. The spectrum is characterized by weak, broad, double-peaked
Balmer and He lines on top of a blue continuum of slope $p \approx
1/3$, as expected for an optically-thick accretion disk. The weak
Bowen blend seen in our spectra may indicate a low metallicity for 
the source.  The presence of a partial S-wave pattern
in the He{\sc {\sc ii}}~$\lambda 4686$ line appears consistent with
the reported photometric period $P_{\rm orb} = 4.1$~hr for
XTE~J1118+480.  By using a combination of Doppler mapping and various
theoretical arguments, we constrain plausible orbital parameters for
the system: a mass ratio $0.02 \lsim q \lsim 0.1$, an inclination $i
\gsim 70^o$ for a neutron star primary, or $30^o \lsim i \lsim 50^o$
for a black hole primary with a mass between $4$ and $10~M_\odot$.
Ca{\sc ii} absorption features observed at very high resolution
constrain the interstellar hydrogen absorption column $\log [N_{\rm
H{\sc i}}~({\rm ~cm^{-2}})] \approx 20.45 \pm 0.2$ and the
identification of three absorbing clouds indicate a distance to the
source $\lsim 1$~kpc, assuming the line-of-sight to XTE~J1118+480 has
average high-latitude properties. These results are discussed in the
context of previous multiwavelength observations of this unusual
system.
\end{abstract}


\section{Introduction}

Soft X-ray Transients (SXTs) are compact binary systems in which a
low-mass secondary (either a main-sequence star or a subgiant)
transfers mass via Roche-lobe overflow onto a black hole (BH) or a
neutron star (NS) primary (see reviews by Tanaka \& Lewin 1995; van
Paradijs \& McClintock 1995; White, Nagase \& Parmar 1995). SXTs have
highly variable luminosities.  They spend most of their lifetime in a
low luminosity quiescent state, but occasionally undergo dramatic
outbursts during which both their optical and X-ray emission increase by
several orders of magnitude (see, e.g., Tanaka \& Shibazaki 1996;
Chen, Shrader \& Livio 1997).

The X-ray emission during outburst of an SXT is typically dominated by
relatively soft, thermal emission from the accretion disk surrounding
the compact object, while the optical emission is usually interpreted
as reprocessed X-rays from the disk and/or the companion star. The new
source XTE~J1118+480 belongs to the class of SXTs, but it also
possesses some rather unusual characteristics.

XTE~J1118+480 was discovered on March 29 2000 with the RXTE All-Sky
Monitor as a brightening X-ray source. Pointed RXTE observations
confirmed the presence of this high Galactic latitude source
($l=157.7^o$, $b=+62.3^o$), with a rather hard power law emission
spectrum of photon index $\Gamma \simeq 1.8$ up to 30~keV. It was
subject to rapid X-ray flares, but no pulsation was
detected. Retrospective ASM analysis revealed that the source
experienced another modest outburst in January 2000 (Remillard et
al. 2000; see Fig.~\ref{fig:lightcurve}). BATSE observations showed
that the source is visible up to 120~keV (Wilson \& McCollough 2000),
and a 6.2 mJy { (at 15 GHz)} variable radio counterpart was later
discovered (Pooley \& Waldram 2000).

The $V \sim 13$ optical counterpart of XTE~J1118+480 was found to
correspond to an $18.8$ mag star in the USNO catalog by Uemura, Kato
\& Yamaoka (2000; see also Uemura et al. 2000).  A photometric
modulation on a 4.1~hr period was observed by Cook et al. (2000) and
was later confirmed as a plausible orbital period by Patterson (2000)
and Uemura et al. (2000; see Stull, Ioannou \& Webb in Haswell et
al. 2000b for a discrepant claim at twice this value). The photometric
modulation was later reported changing shape and period, possibly
showing the development of superhumps in this source (Uemura 2000).
{ If these are normal superhumps, then the orbital period would be
smaller than the photometric superhump modulation by a few percent.}

Garcia et al. (2000) reported the first optical spectroscopic results
on XTE~J1118+480. They found an optical spectrum typical of SXTs in
outburst, with very broad H$\alpha$, H$\beta$ and He{\sc ii} lines (
FWHM $\gsim 2000$~km~s$^{-1}$). These observations also indicated the
presence of absorption features and a very low interstellar absorption
($E(B-V) \lsim 0.024$) to the source. These authors suggested that the
surprisingly low X-ray ($\sim 40$~mCrab, 2-12~keV) to optical ($V \sim
13$) flux ratio of XTE~J1118+480 could be due to a nearly edge-on
viewing angle.

Additional X-ray observations revealed the presence of a strong
Quasi-Periodic Oscillation (QPO) at 0.085~Hz in the X-ray lightcurve
of XTE~J1118+480. The shape of the power density spectrum and the hard
emission spectrum of XTE~J1118+480 prompted Revnivtsev, Sunyaev \&
Borozdin (2000) to propose that the source is a BH transient, by
analogy with other such systems. The QPO was confirmed by ASCA
observations, which also suggest the presence of a soft component in
the spectrum (below 2~keV), possibly due to emission from the
accretion disk in the system (Yamaoka et al. 2000). { Soft
X-ray observations by {\it Chandra} did not confirm this
(McClintock et al. 2000)}. The QPO frequency
was reported to have shifted from 0.085~Hz to $\gsim 0.1$~Hz in
subsequent observations (Yamaoka et al. 2000; Wood et al. 2000).
Pointed XTE observations at the end of May do not reveal significant
changes in the emission spectrum (P. Jonker, private communication).

XTE~J1118+480 has also been the subject of an extensive
multiwavelength observation campaign with HST, EUVE, UKIRT and
RXTE. Thanks to the very low interstellar absorption to the source,
the first EUVE spectrum of a BH-candidate X-ray Transient was obtained
(Mauche et al. 2000). No periodic modulation was found in the EUVE
data (Hynes et al. 2000b).  An HST spectrum revealed a very broad ($>
10,000$~km~s$^{-1}$) Ly$\alpha$ absorption feature, suggestive of a
massive accretor (Haswell et al. 2000b). Haswell et al. (2000c)
obtained a near-UV power density spectrum of XTE~J1118+480 with a QPO
and an overall shape in agreement with previous RXTE timing data. In
addition, the near-UV variability was found to lag by 1-2~seconds
behind the X-ray variations, as would be expected from light echoes in
a system with $P_{\rm orb} =4.1$~hr. Hynes et al. (2000b), combining
data from HST, EUVE, UKIRT and RXTE, find that the IR to UV data
suggests emission from both an optically-thick disk and another, flat
spectrum component (possibly synchrotron), while the EUVE and X-ray
data suggest a power law emission typical of a Galactic X-ray binary
in a low-hard state. They conclude that XTE~J1118+480, rather than
experiencing a full outburst { approaching the Eddington
luminosity}, seems to be in a mini-outburst state { (but see
Kuulkers 2000)}.

In this paper, we describe the results of our optical spectroscopic
campaign to monitor the evolution of XTE~J1118+480 during its recent
outburst (and early decline). In \S2, we describe our
observations and the data reduction techniques used. Our results
concerning the continuum and line emission, a partial S-wave pattern
and the interstellar absorption to the source are presented in \S3 and
discussed in the context of the current knowledge on XTE~J1118+480 in
\S4. Our main conclusions are summarized in \S5.

\section{Observations and Reduction}

We obtained optical spectra of XTE~J1118+480 from April 7, 2000 to
July 4, 2000 with the ARC 3.5~m telescope at Apache Point Observatory.
{ We mostly used the Double Imaging Spectrograph (DIS),} but
spectra with the Echelle spectrograph were also obtained on April 7,
2000. Table~\ref{tab:obs} summarizes the dates and other
characteristics of our observational campaign.

Most of our observations with DIS were carried out with the high
resolution gratings { (hereafter {\em hires})} with a 1.5'' slit
(dispersion 1.6\AA~pixel$^{-1}$ in the blue, 1.1\AA~pixel$^{-1}$ in
the red, and a resolution of 2~pixels), but we also obtained spectra
with the low-resolution gratings (same slit size, dispersion
6.2\AA~pixel$^{-1}$ in the blue, 7\AA~pixel$^{-1}$ in the red, and a
resolution of 2~pixels;{ hereafter {\em lowres})} on April 7,
2000. For the DIS high resolution observations, the blue and red
gratings were centered on slightly different wavelengths during our
various nights of observations, but we generally centered the blue
side to cover the H$\beta$ and He{\sc ii}~$\lambda 4686$ lines, and
the red side to cover the H$\alpha$ line. The complete list of
spectral coverage for our observations can be found in
Table~\ref{tab:obs}.  The Echelle spectrograph covers the (fixed)
wavelength range 3500-10000\AA\ with $R\sim 30000$ (10~km~s$^{-1}$ at
5000\AA) and a resolution element of $\sim$~2.5 pixels.  All the
exposure times can be found in Table~\ref{tab:obs}.

The DIS observations were reduced in the standard way using {\em IRAF}
and the spectra were optimally-extracted and dispersion corrected
without any particular difficulty.

The Echelle spectra were reduced using the IRAF {\em ecspec} package.
Direct extraction of the object spectra proved difficult for the
highest (bluest) orders where the trace was hard to follow. We decided
to use the flat field as a guide and validated the method with a
standard star observed on the same night.  This resulted in
significantly higher S/N in the blue part of the spectrum. A total of
115 orders were extracted covering the spectral range 
3500\AA-10000\AA, in which the source is detected, albeit with
varying sensitivity.  The dispersion varied between
$\sim$~0.05\AA/pixel to 0.1\AA/pixel and the usable spectral range in
each order between $\sim$ 70\AA\ to 150\AA\ from the blue to the red
end.

\begin{figure}
\plottwo{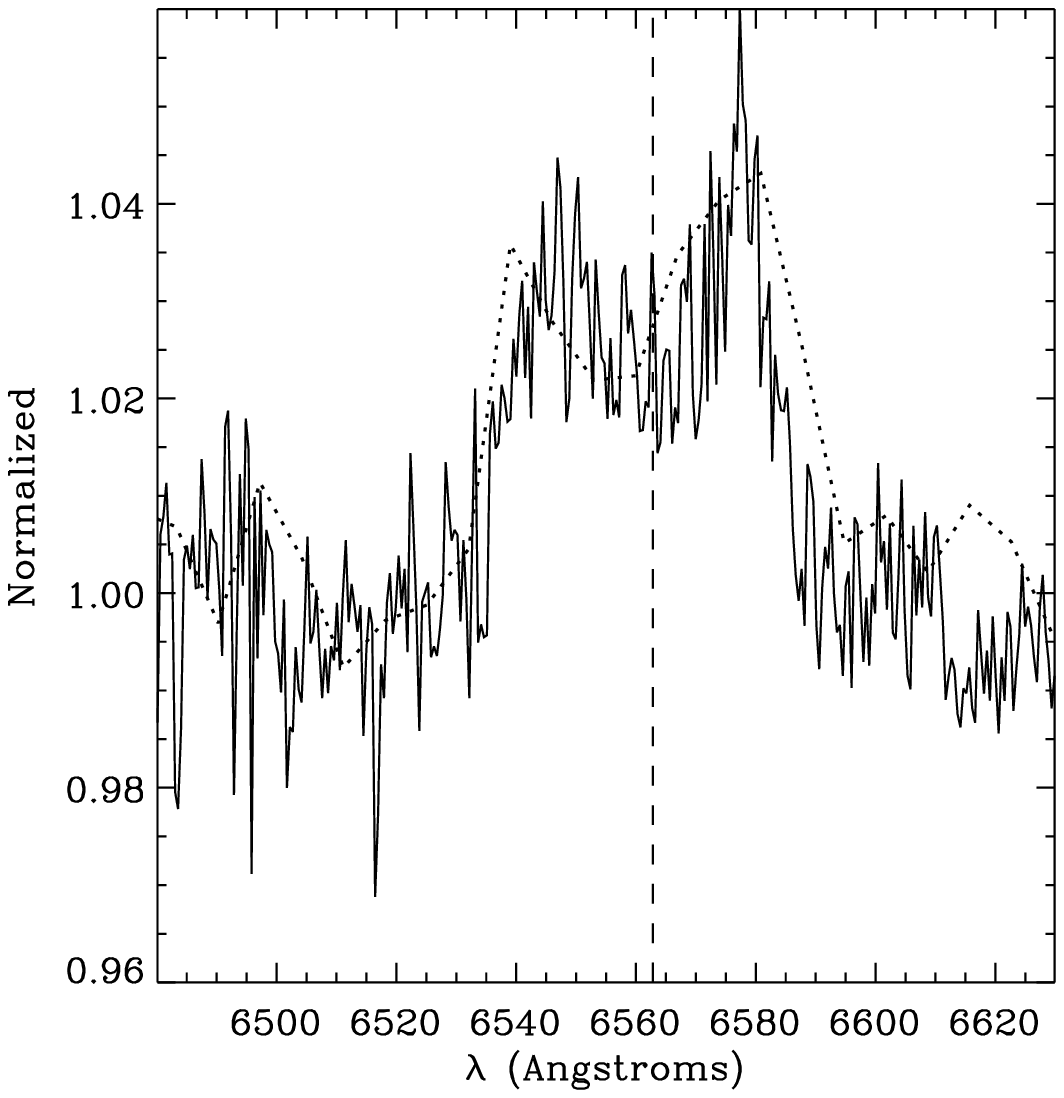}{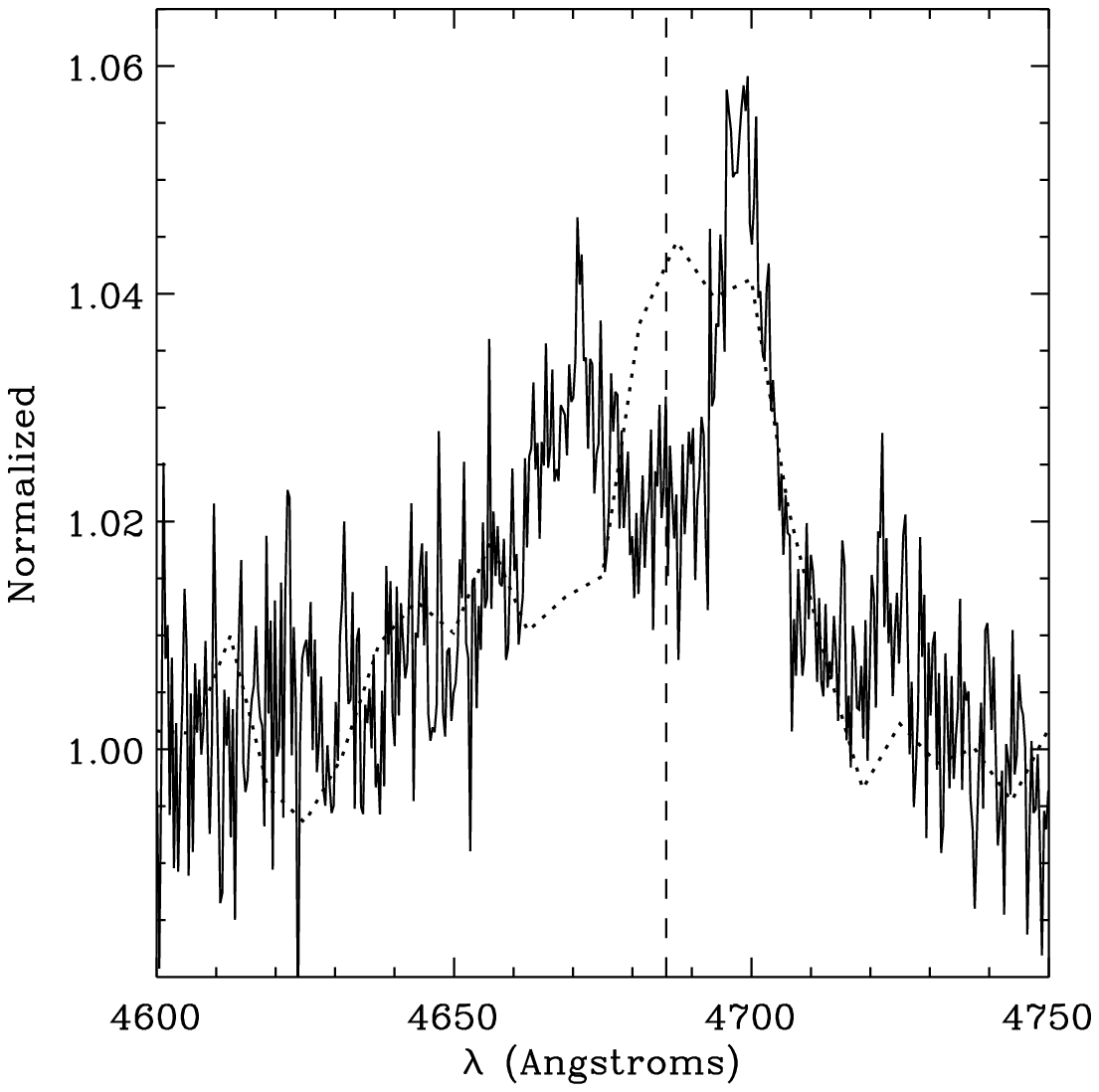}
\caption{H$\alpha$ (left) and He{\sc ii} (right) profiles for the summed Echelle spectra of April 7. Overplotted are the low resolution spectra taken earlier on the same night. The Echelle spectrum were also binned to
0.50\AA/pix for H$\alpha$ and 0.35\AA/pix for He{\sc ii}.
\label{fig:echha}}
\end{figure}

The emission lines in the spectrum of XTE~J1118+480 are both broad and
weak (see below, Fig.~\ref{fig:lowres}) and their identification in
the Echelle spectra turned out to be challenging. The broad lines can
cover almost half of the spectral range in one order making a
continuum fit unreliable.  We therefore proceeded by normalizing to an
interpolated continuum determined from the two closest orders where no
lines are expected. The normalised summed spectra clearly show
H$\alpha$ (in one order, left panel of Fig.~\ref{fig:echha}) and
He{\sc ii} $\lambda$4686 (spread over two orders, right panel of
Fig.~\ref{fig:echha}). These lines are also detectable in the
individual continuum-subtracted { Echelle} spectra. { The
different profile of the He{\sc ii} line in the {\em lowres} and
Echelle spectra is most probably due to the S-wave modulation
discussed below. This S-wave is much less prominent in H$\alpha$}.

\section{Results}

\begin{figure}
\plotone{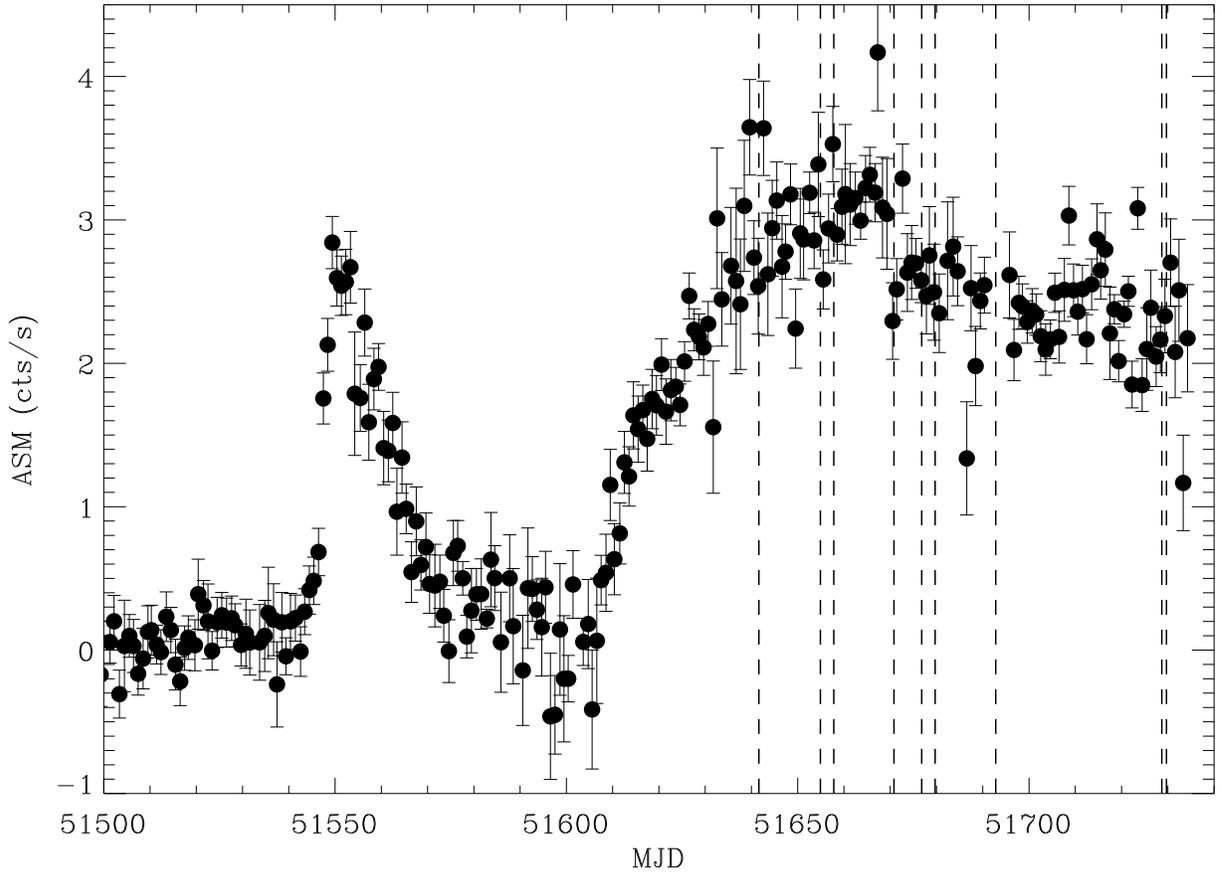}
\caption{RXTE ASM lightcurve of XTE~J1118+480, with dashed lines
indicating the dates of our spectroscopic observations (see
Tab.~\ref{tab:obs}). Data provided by the ASM/RXTE teams at MIT and at
the RXTE SOF and GOF at NASA's GSFC.
\label{fig:lightcurve}}
\end{figure}

The RXTE ASM lightcurve of XTE~J1118+480 is shown in
Fig.~\ref{fig:lightcurve} with the dates of our spectroscopic
observations indicated by dashed lines.  The source was in outburst
during the entire period covered by our optical observations, with a
flux about 40~mCrab in the 2-12~keV band
(Fig.~\ref{fig:lightcurve}). Optical photometry
\footnote{VSNET observations at
http://www.kusastro.kyoto-u.ac.jp/vsnet} also shows the source in
outburst with a mean magnitude that decreased at most by $0.5$~mag
(from its peak value $\lsim 13$) at the end of our observational
program. Since then, the optical and X-ray flux have decreased
significantly, indicating that the source may soon enter quiescence.

\subsection{Continuum}

\begin{figure}
\plotone{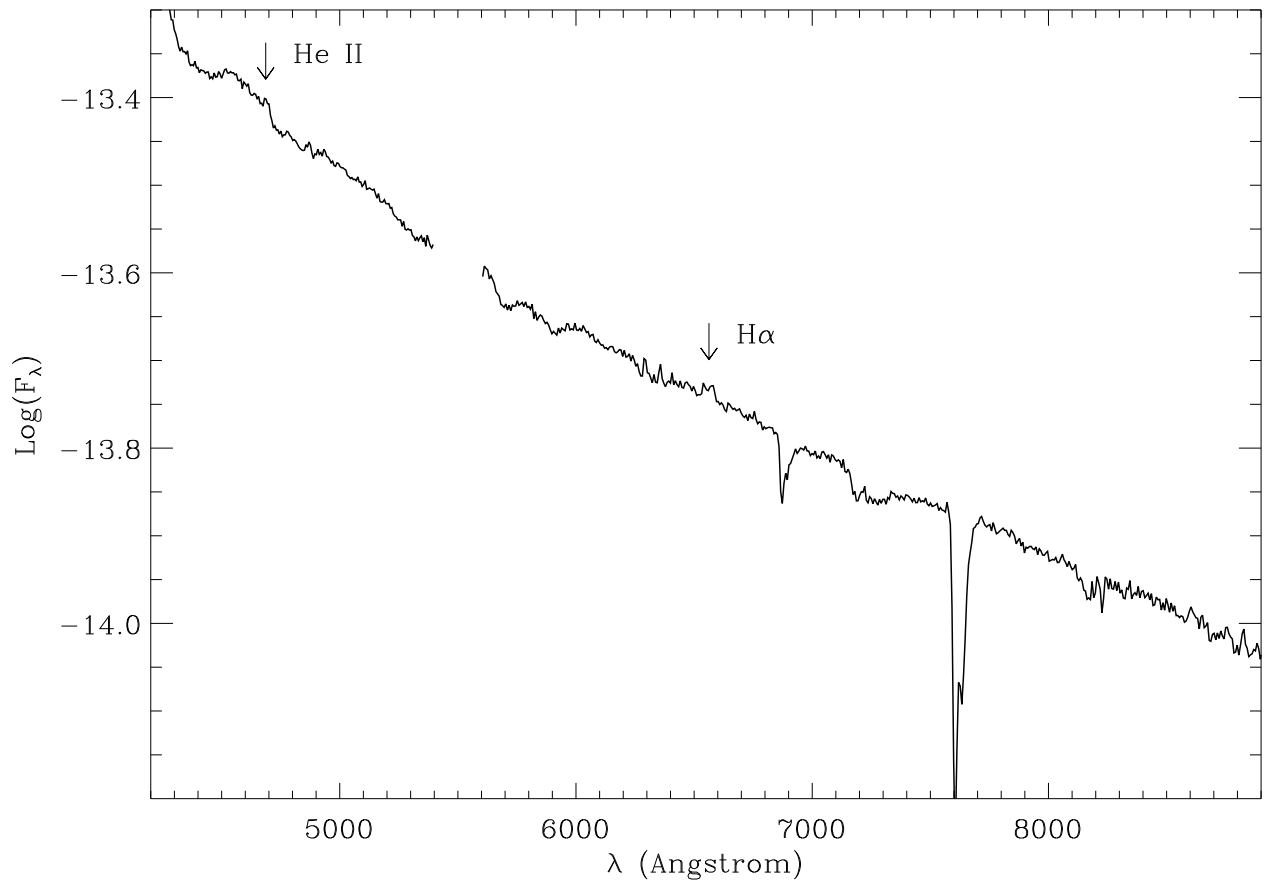}
\caption{A representative spectrum of XTE~J1118+480 obtained by
summing the two DIS {\em lowres} spectra of April 7, 2000. The spectrum,
typical of an X-ray transient in outburst, is characterized by a blue
power-law continuum and the presence of relatively weak double-peaked
emission lines, such as H$\alpha$ and He{\sc ii}~$\lambda 4686$. The blank
spectral region corresponds to wavelengths which are not efficiently
covered by either the blue or the red side of the spectrograph. The
unit for F$_\lambda$ is erg~s$^{-1}$~cm$^{-2}$~$\AA ^{-1}$.
\label{fig:lowres}}
\end{figure}

Figure~\ref{fig:lowres} shows the sum of the two DIS {\em lowres} spectra
obtained on April 7, 2000, which is also representative of other
spectra of the source obtained later on during our campaign.  The
strong, blue power-law continuum and the weak emission lines are
typical of an SXT in outburst (see also Garcia et al. 2000). A
power-law fit to the blue part of the spectrum yields a slope $p
\simeq +0.4 \pm 0.2$ (where the flux F$_\nu \propto \nu^p$), while a
similar fit to the red part of the spectrum yields $p \simeq +0.33 \pm
0.15$. These values are consistent with the expectation $p=1/3$ for an
optically-thick accretion disk (see, e.g., Frank, King \& Raine
1992). Hynes et al. (2000b) find that the spectrum is flat using a
wider wavelength range.

The broad, double-peaked H$\alpha$ emission line is the only one
easily identified on the spectrum of Fig.~\ref{fig:lowres}, but a
closer examination reveals the presence of He{\sc ii}~$\lambda4686$,
and of additional Balmer and He lines in the other spectra that we
collected (see \S3.2 below). We find only marginal evidence for
continuum variability in XTE~J1118+480, both during the same night and
from night to night. The lack of obvious variability is consistent
with the long-term variations in the optical VSNET lightcurve ($<$
0.5~mag), the short timescale ($<$10~s) of the large amplitude
($>$0.3~mag) flickering and the low amplitude (0.052 mag) of the 4~hr
modulation (Patterson 2000).

\subsection{Line Profiles and their Evolution during Outburst}

We identify three Balmer lines, three He lines and the Bowen blend in
our set of spectra.  These are: H$\gamma~\lambda 4340.5$, He{\sc
i}~$\lambda 4471.5$, the Bowen blend at $\sim~\lambda4638$, He{\sc
ii}~$\lambda 4686.7$, H$\beta~\lambda 4861.3$, He{\sc i}~$\lambda
4921.9$, H$\alpha~\lambda 6562.8$ and He{\sc i}~$\lambda 6678.1$ The
lines are weak, broad and double-peaked. H$\beta$, H$\gamma$ and also
the weaker He{\sc i} $\lambda$4471.5 have double-peaked emission
clearly embedded in a large absorption trough. There is no clear
evidence of such absorption around H$\alpha$.

\begin{figure}
\plottwo{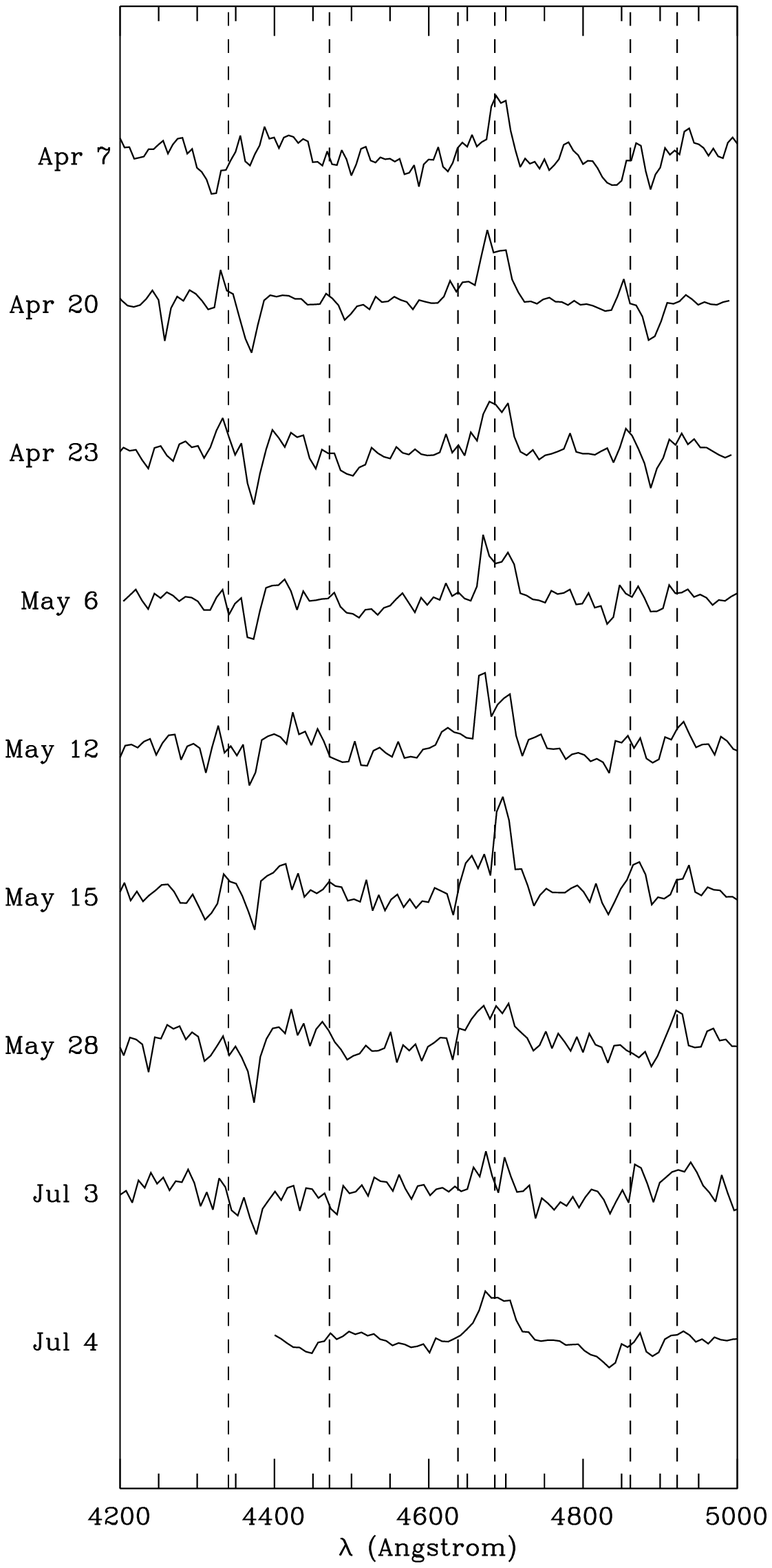}{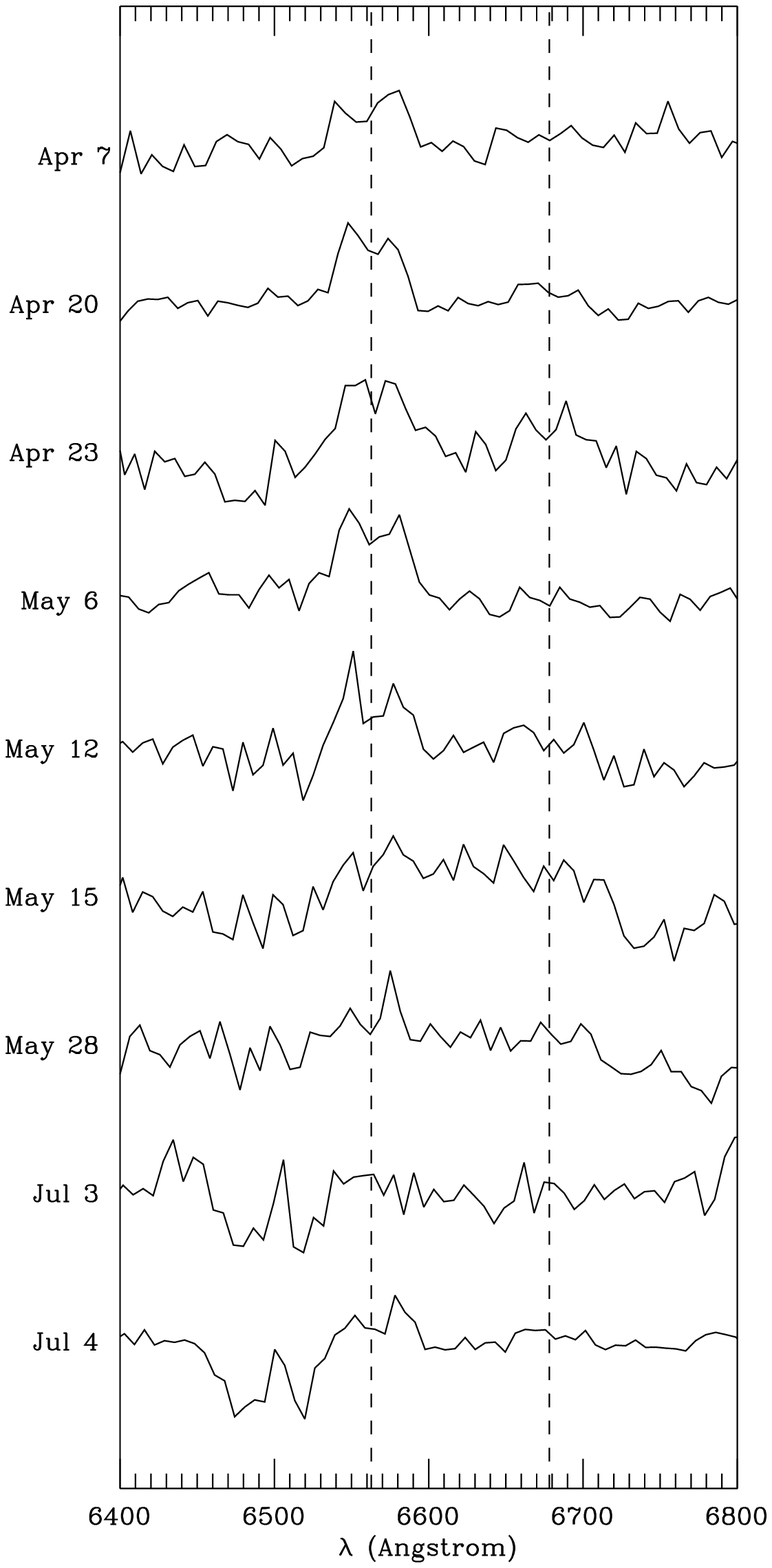}
\caption{Evolution of the line profiles over the 3-month period of
observation. The location of H$\gamma~\lambda 4340.5$, He{\sc
i}~$\lambda 4471.5$, the Bowen blend at $\sim~\lambda4638$, He{\sc
ii}~$\lambda 4686.7$, H$\beta~\lambda 4861.3$, He{\sc i}~$\lambda
4921.9$, H$\alpha~\lambda 6562.8$ and He{\sc i}~$\lambda 6678.1$ are
indicated by vertical dashed lines. The spectra are the { daily}
sums of the DIS {\em hires} spectra except for the top one which is
the sum of the two DIS {\em lowres} spectra obtained on April 7,
2000. The DIS {\em hires} spectra were further binned to reach a
resolution comparable to that of the DIS {\em lowres} spectra. The
vertical scale is identical for all spectra which were simply offset
from each other. Note the two unidentified absorption features
blueward of H$\alpha$.
\label{fig:sumlines}}
\end{figure}

Fig.~\ref{fig:echha} shows the H$\alpha$ and He{\sc ii} line profiles
in the summed Echelle spectrum with the low resolution spectrum taken
on the same night overplotted. We find no evidence for structures at
high spectral resolution in the lines, except for an excess at rest
wavelength which would not be expected in a pure double-peaked disk
profile. This component also seems to be present in the high
resolution DIS spectra (see Fig.~\ref{fig:sumlines}). The Bowen blend,
which is very weak in the DIS {\em hires} spectra, was not detected on April
7 in either the DIS {\em lowres} or the Echelle spectra.

Measuring the lines proved difficult, not because of a lack of counts
in the spectra but because of their intrinsic broadness and
weakness. As a rule, we performed two-component Gaussian fits to the
emission lines. For H$\gamma$ and H$\beta$ the two emission components
were subtracted to measure the equivalent width (EW) and FWHM of the
absorption trough. The EW of the lines varies (with large associated
errors) within a night but we could find no periodic behaviour linked
to the { photometric} period. We also tried to estimate the semi-amplitude
velocity of the primary $K_1$ by fitting the wings of the He{\sc ii}
line (expected to trace the motion of the compact star) and folding
around the suspected $P_{\rm orb}$ of 0.17~day. This did not lead to
any result as could be expected from the usually low $K_1$ of SXTs
($K_1\lsim 50$~km~s$^{-1}$).

The evolution of these lines during the 3 months of observations is
shown in Fig.~\ref{fig:sumlines}, after the underlying continuum was
normalized to unity. When available, multiple spectra were summed up
to increase the S/N of the spectra.  A comparison between the most
comprehensive datasets (Apr. 17 and Jul. 4), where measurement errors
can be minimized, shows that the two strongest lines, H$\alpha$ and
He{\sc ii}, did not change significantly. We conclude that there do
not seem to be any long term variations.  The average properties of
the lines are summed up in Table.~\ref{tab:lines} (the quoted errors are
the rms of the measurements).

In all our spectra the blue side of the absorption troughs in
H$\gamma$ and H$\beta$ is less { strong} than the red side. The
measured central wavelengths of the absorptions are redshifted
compared to the rest wavelength (which could be an artefact of the
method we used to remove the double peaked emission). Fitting a
gaussian to the red part of the absorption while ignoring (but not
subtracting) the emission lines gave higher equivalent widths (2.4\AA\
for H$\gamma$ and 2.0\AA\ for H$\beta$) and better agreement with the
rest wavelength (7\AA\ redshift for both instead of 15\AA). Similar
absorption redshifts were reported by Callanan et al. (1995) for
GRO~J0422+32. A plausible explanation would be that the shifts are due
to distorted line profiles. Such asymetric lines are observed in dwarf
novae when the disk becomes eccentric and the system shows superhumps
(Warner 1995). There is evidence for such superhumps in XTE~J1118+480
(see discussion in \S4.3).

\subsection{Constraint on the EUV flux}

The He{\sc ii} emission may be used to obtain a crude estimate of the
extreme ultraviolet (EUV) flux from the source (e.g. Patterson \&
Raymond 1985) { if one assumes} the $\lambda$4686 line stems from
the recombination of He{\sc i} photoionized in the disk by photons
with energies between 55-280 eV. { On the other hand, the doppler
map discussed below in \S3.5 shows most of the He{\sc ii} emission is
localized and in all likelihood associated with stream-disk
interaction. In this case the $\lambda$4686 line would be pumped by
collisional excitation rather than photoionization. The following
estimate therefore only yields an upper limit on the EUV flux. This is
still of interest since the upper limit is independent of the column
density to the source.}

We derive from our observations $F_{\lambda4686}\approx
7\cdot10^{-14}$ erg~s$^{-1}$~cm$^{-2}$ and, from the values given by
Haswell et al. (2000b), $F_{\lambda1640}\approx 7\cdot10^{-13}$
erg~s$^{-1}$~cm$^{-2}$.  This favours Case B recombination for which
one would expect $F_{\lambda1640}/F_{\lambda4686}\sim7$ (Seaton 1978).
Therefore, the region is optically-thick to the ionizing flux but thin
to the line and a fraction $\epsilon\approx0.2$ of the photoionized
He{\sc i} recombinations lead to $\lambda$4686 emission (Hummer \&
Storey 1987).

Following previous applications to soft X-ray transients (Marsh,
Robinson \& Wood 1994; Hynes et al. 1998), the EUV flux $F_{\rm EUV}$
and the He{\sc ii} line flux $F_{\lambda4686}$ are related through
\begin{equation}
\epsilon\alpha\frac{F_{\rm EUV}}{E_{\rm EUV}}=\frac{F_{\lambda4686}}
{E_{\lambda4686}}
\end{equation}
where $\alpha$ is the fraction of EUV photons intercepted by the disc
while $E_{\rm EUV}\approx100$~eV and $E_{\lambda4686}\approx2.6$~eV
refer to the mean energy of the photons.  We have assumed $L_{\rm
EUV}/L_{\lambda4686}=F_{\rm EUV}/F_{\lambda4686}$, which may be
incorrect if the EUV emission is not isotropic or if we see only a
fraction of the He{\sc ii} emission. In both cases this assumption
leads to an underestimate of the true EUV luminosity. Observations
suggest that a fraction $\sim 10^{-3}$ of the soft X-ray flux is
intercepted and reprocessed in the optical by the accretion disk in
SXTs (e.g. Dubus et al. 1999). By analogy, we take $\alpha=10^{-3}$
from which we finally derive { an upper limit to the} 55-280~eV
flux of $10^{-8}$ erg~s$^{-1}$~cm$^{-2}$. This is obviously a very
crude estimate.

{ An EUVE spectrum was obtained for this source, the first time
ever for a SXT. The source flux at these wavelengths is heavily
dependent on the extinction, but the upper limit on $F_{\rm EUV}$
derived above appears consistent with both a simple extrapolation of
the optical--UV flux observed and the EUVE flux inferred if a value of
$N_{\rm H} \approx 10^{20}$~cm$^{-2}$ is assumed when dereddening the
EUVE spectrum (Mauche et al. 2000; Hynes et al. 2000).  }

\subsection{Peak-to-peak velocities}

The mean peak-to-peak separation for all the observations is 18\AA\
(1240 km~s$^{-1}$) for H$\gamma$, 23\AA\ (1470 km~s$^{-1}$) for He{\sc
ii}, 21\AA\ (1300 km~s$^{-1}$) for H$\beta$ and 28\AA\ (1280
km~s$^{-1}$) for H$\alpha$. The S-wave is much less prominent in the
Balmer lines { and} we assume this represents the projected
Keplerian velocity of the outer disk.  { The peak-to-peak
separations have been found in well-known systems to overestimate the
outer disk radius by about 20\% because of sub-Keplerian motion or
local broadening (Marsh 1998). This uncertainty is acceptable
considering other assumptions made below.} For a Keplerian disk with
an emissivity $\propto R^{-n}$, $\Delta v$ can be related to the outer
(emitting) disk radius $R_{\rm d}$ (Smak 1981):
\begin{equation}
\Delta v = 2 R_{\rm d}\Omega_{\rm K} \sin i = 2 \sin i (GM_1/R_{\rm
d})^{1/2},
\label{eq:vrout}
\end{equation}
where $M_1$ is the mass of the primary and $i$ the system inclination.
We can also express $R_{\rm d}/a$ as a function of the
velocity semi-amplitude of the secondary $K_2$ and the mass ratio
$q=M_2/M_1$ by combining the above equation with the mass function
(see below, Eq.~\ref{eq:fm}) and Kepler's third law\footnote{From
$P_{\rm orb}=$0.17~day, Kepler's third law gives a binary separation
$a \approx 10^{11}$~cm with a weak dependence on $M_1$ and $q$.},
resulting in:
\begin{equation}
\frac{R_{\rm d}}{a}=(1+q)\left(\frac{2 K_2}{\Delta v}\right)^2
\label{eq:rova}
\end{equation}
The disk outer radius is expected to reach the tidal truncation radius
for $q\gsim0.2$ and to be limited to $R_{\rm d}/a\approx0.48$ by the
3:1 resonance for $q\lsim 0.2$ (e.g. Warner 1995). Papaloizou \&
Pringle (1977; their Tab.~1) give a theoretical estimate of the outer
disk radius that we recklessly use for $R_{\rm d}/a$.
Eq.~\ref{eq:rova} can therefore be used to give the semi-amplitudes
$K_2(q)$ for which the disk radius is consistent with the theoretical
expectations.  For $0.01<q<1$ the disk radius varies between 0.5 and
0.3 in units of the binary separation $a$. The expected $K_2$ as a
function of $q$ is shown as a solid line in the left panel of
Fig.~\ref{fig:cons1}.  We find $440>K_2$~(km~s$^{-1}$)~$>260$ for
$0.01<q<1$.

The observed velocity FWHM ($v_{\rm fwhm}$) of the lines provides an
upper limit to the minimum emitting Keplerian disk radius:
\begin{equation}
R_{\rm in}\lsim 1.3\cdot 10^{10}\ (M_1/M_\odot)\left(\sin i/v_{\rm
fwhm}\right)^2\ {\rm cm},
\end{equation}
where $v_{\rm fwhm}$ is expressed in units of 1000~km~s$^{-1}$. The
absorption features from the optically-thick disk (see \S4.1 and
Table.~\ref{tab:lines}) have FWHM $\gsim 3000$~km~s$^{-1}$ implying
$R_{\rm in}\lsim 10^9$~cm.  Haswell et al. (2000b) detect broad
Ly$\alpha$ absorption with $v_{\rm fwhm} \sim 10^4$~km~s$^{-1}$ which
gives a stricter $R_{\rm in}\lsim 10^8$~cm or $R_{\rm in} \lsim 500$
in Schwarzschild units.  Spectral models of the X-ray low/hard state
of BH candidates tend to predict larger values of $R_{\rm in} (\gsim
1000$ in Schwarzschild units) at which the transition from a thin disk
to a hot advection-dominated flow occurs (see, e.g., Esin et
al. 1998). The EUV flux of XTE~J1118+480 could also imply a smaller
transition radius (see \S4.2).

\subsection{Doppler mapping}

\begin{figure}
\plottwo{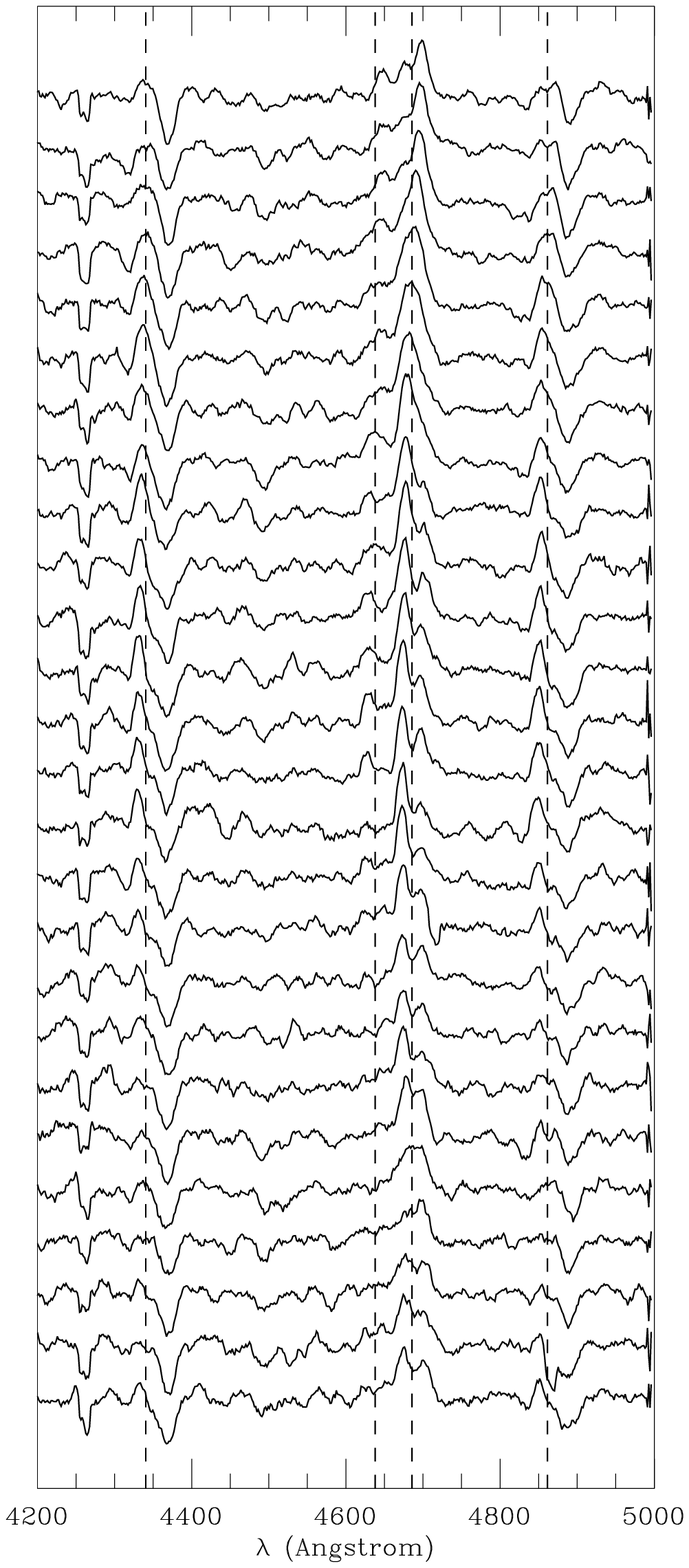}{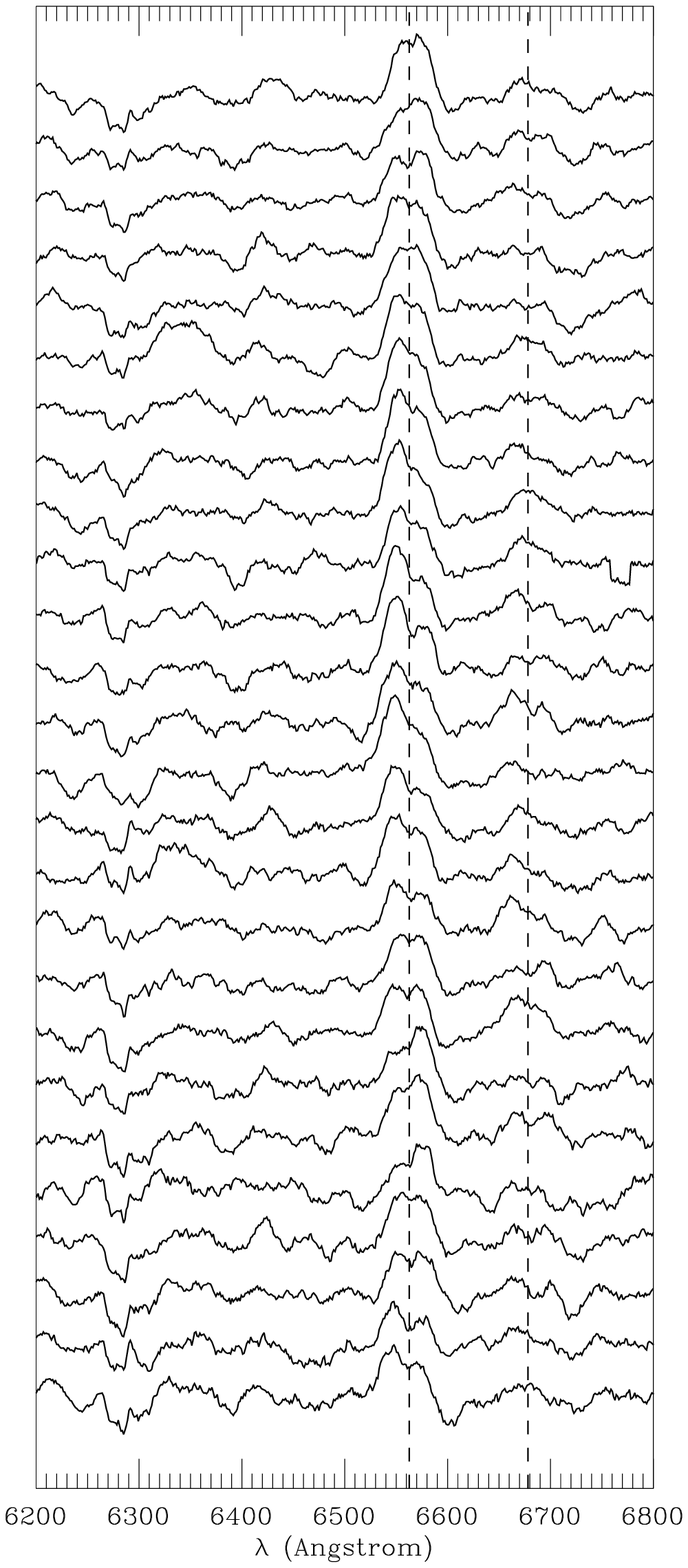}
\caption{Time-resolved spectroscopy of XTE~J1118+480 on April 20, 2000
reveals the presence of an S-wave pattern in the He{\sc ii}~$\lambda
4686$ emission as well as in the Bowen blend, H$\beta$ and H$\gamma$\
(dashed lines).  The S-wave is less prominent in H$\alpha$.  The 26
spectra (which have been smoothed to enhance the features) have
5~minute exposures and the set covers a total of 2.9~hours. The S-wave
pattern, which is moving from the red to the blue side of the
double-peaked line, is consistent with the claimed photometric
orbital period of 4.1~hr.\label{fig:swave}}
\end{figure}

Figure~\ref{fig:swave} shows the evolution of the He{\sc ii}~$\lambda
4686$ line profile over an approximately continuous 2.9 hr period on
April 20, 2000. A partial S-wave pattern moves from the red to the
blue side of the He{\sc ii}~$\lambda 4686$ line rest wavelength and
appears consistent with the claimed photometric period of $4.1$~hr
(Cook et al. 2000; Patterson 2000). The modulated component is
particularly strong towards the middle of the observation at HJD
51654.876 where it is blueshifted by about 900 km~s$^{-1}$.

The trailed spectrum clearly shows that the Bowen blend, He{\sc i}
$\lambda4921.9$, H$\gamma$ and H$\beta$ follow the same S-wave pattern
(see Fig.~\ref{fig:swave} ). We note that identical behaviour is seen
in the smaller continuous set (1.8 hr) of July 4 although with poorer
S/N (not shown here). However, the presence of the S-wave in the
H$\alpha$ emission profile is not clear on either dates.

\begin{figure}
\plottwo{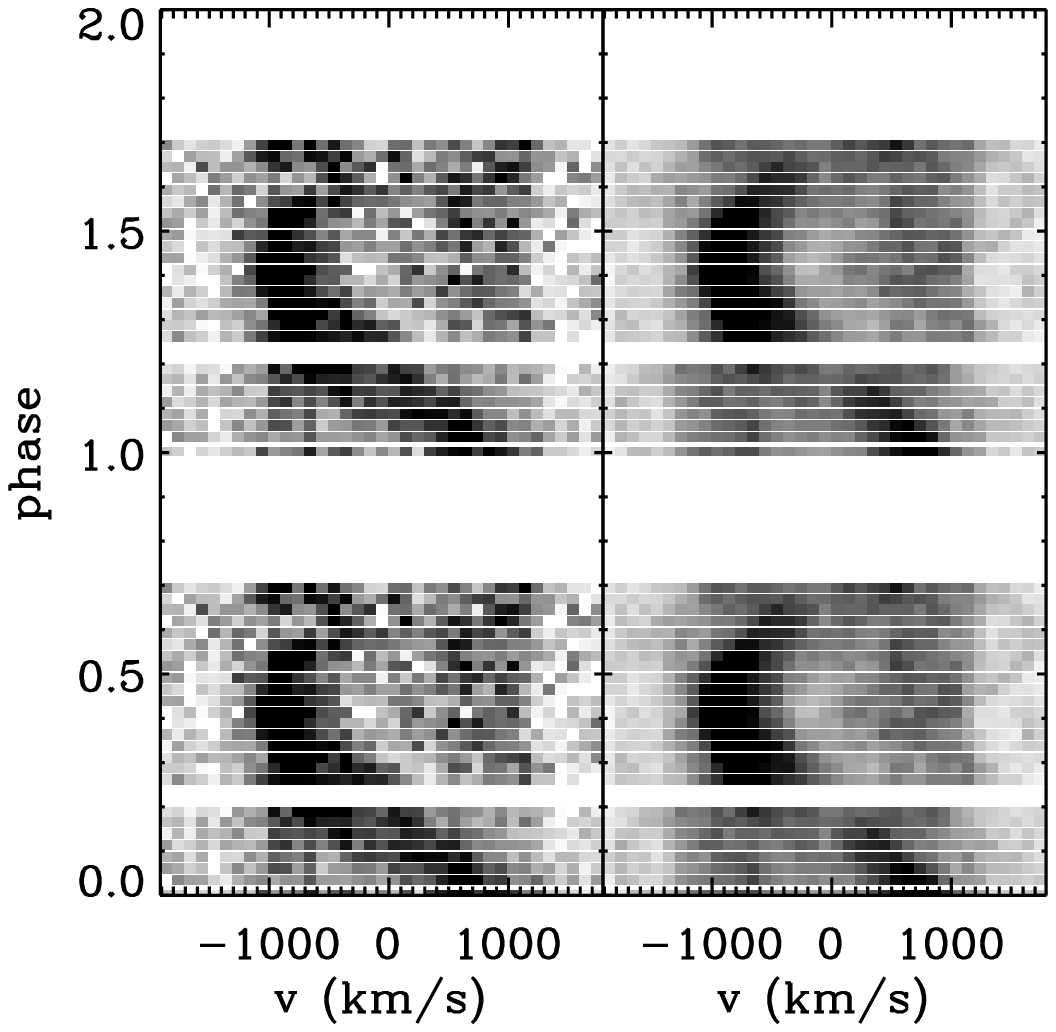}{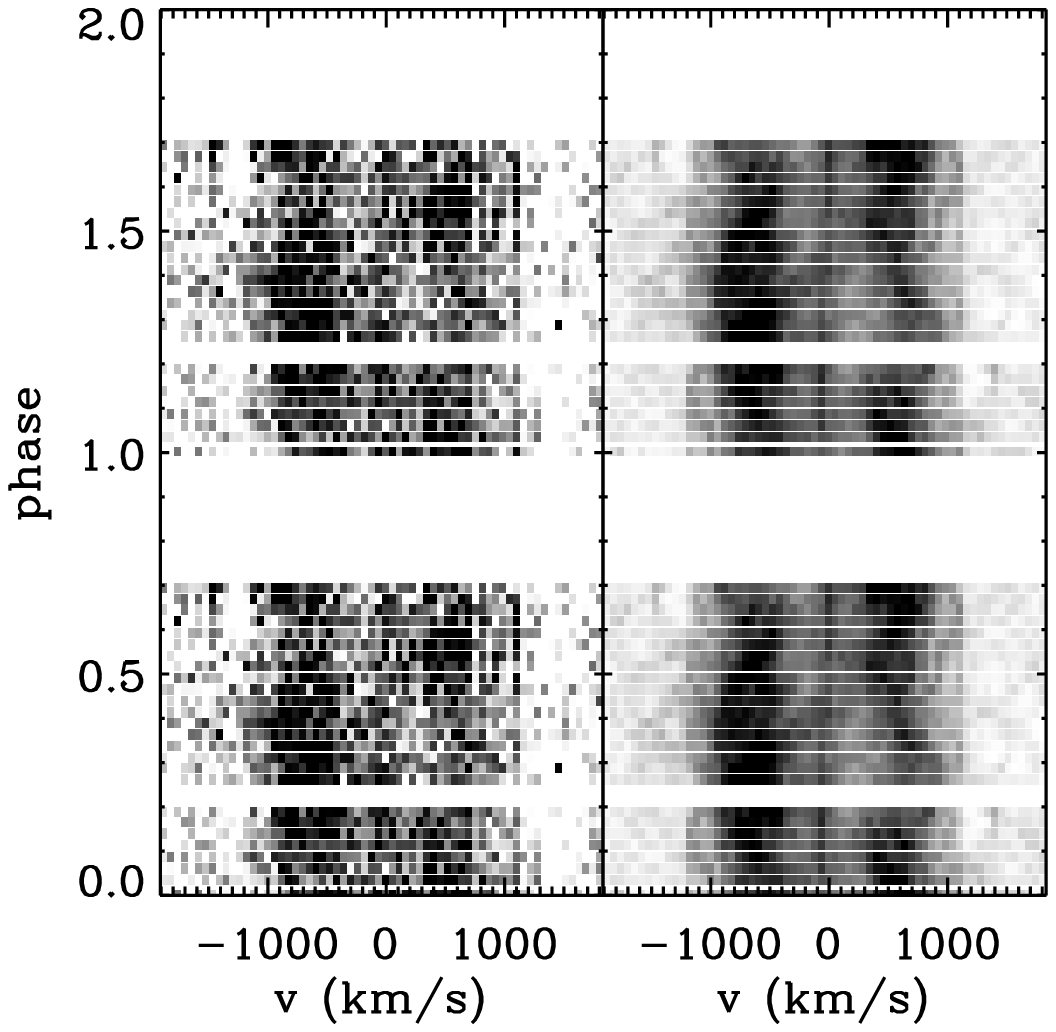}
\caption{Observed (unsmoothed data) and reconstructed trailed emission
line profiles on April 20, 2000 of He{\sc ii} $\lambda4686$ (left) and
H$\alpha$ (right). The corresponding Doppler maps are shown in
Fig.~\ref{fig:map}. The assumed orbital period is 0.1708~day, the
system has a null systemic velocity and the orbital phase is
arbitrary. The velocity reference is the rest wavelength of the
emission line. Note the clear S-wave in the He{\sc ii} emission
profile. { The data are shown twice for clarity.}\label{fig:prof}}
\end{figure}

\begin{figure}
\plottwo{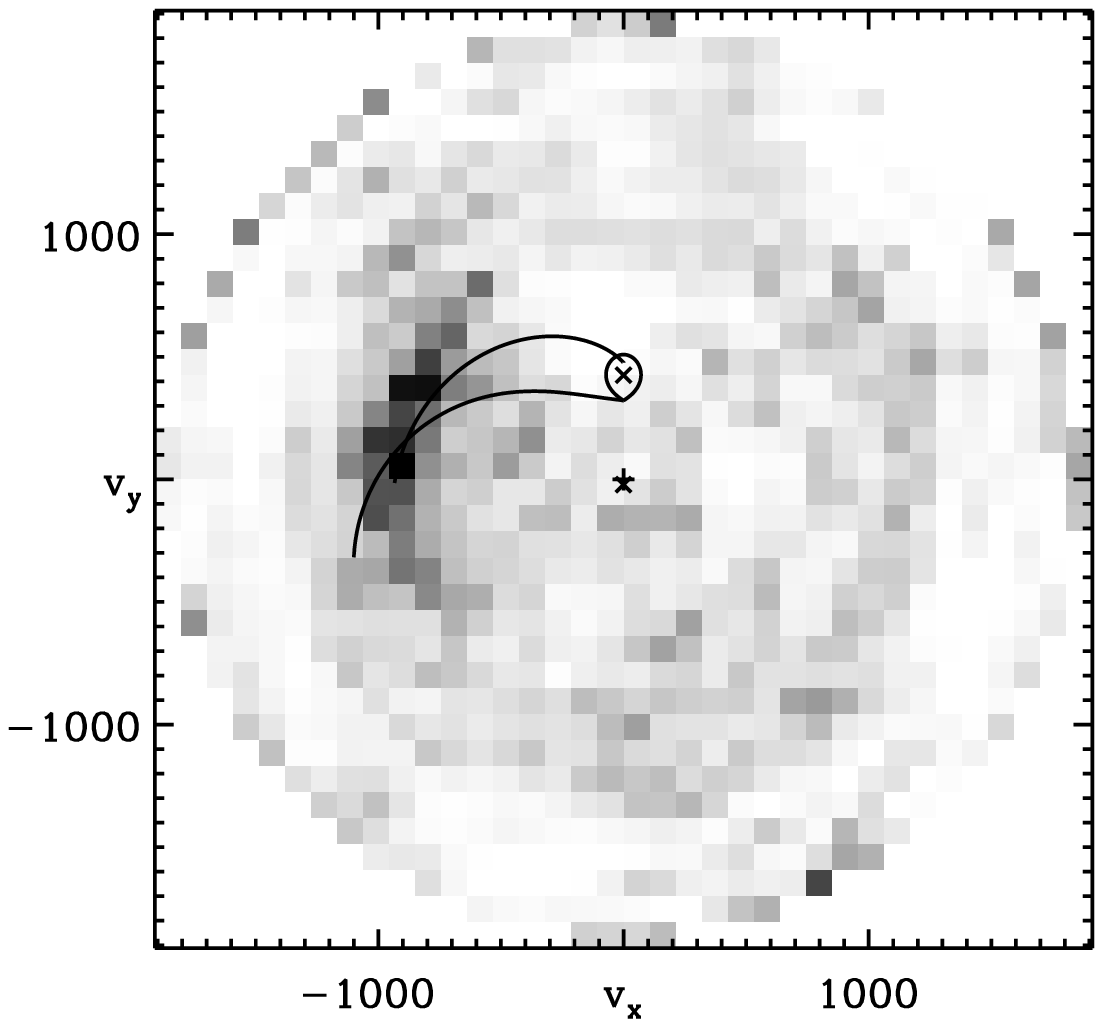}{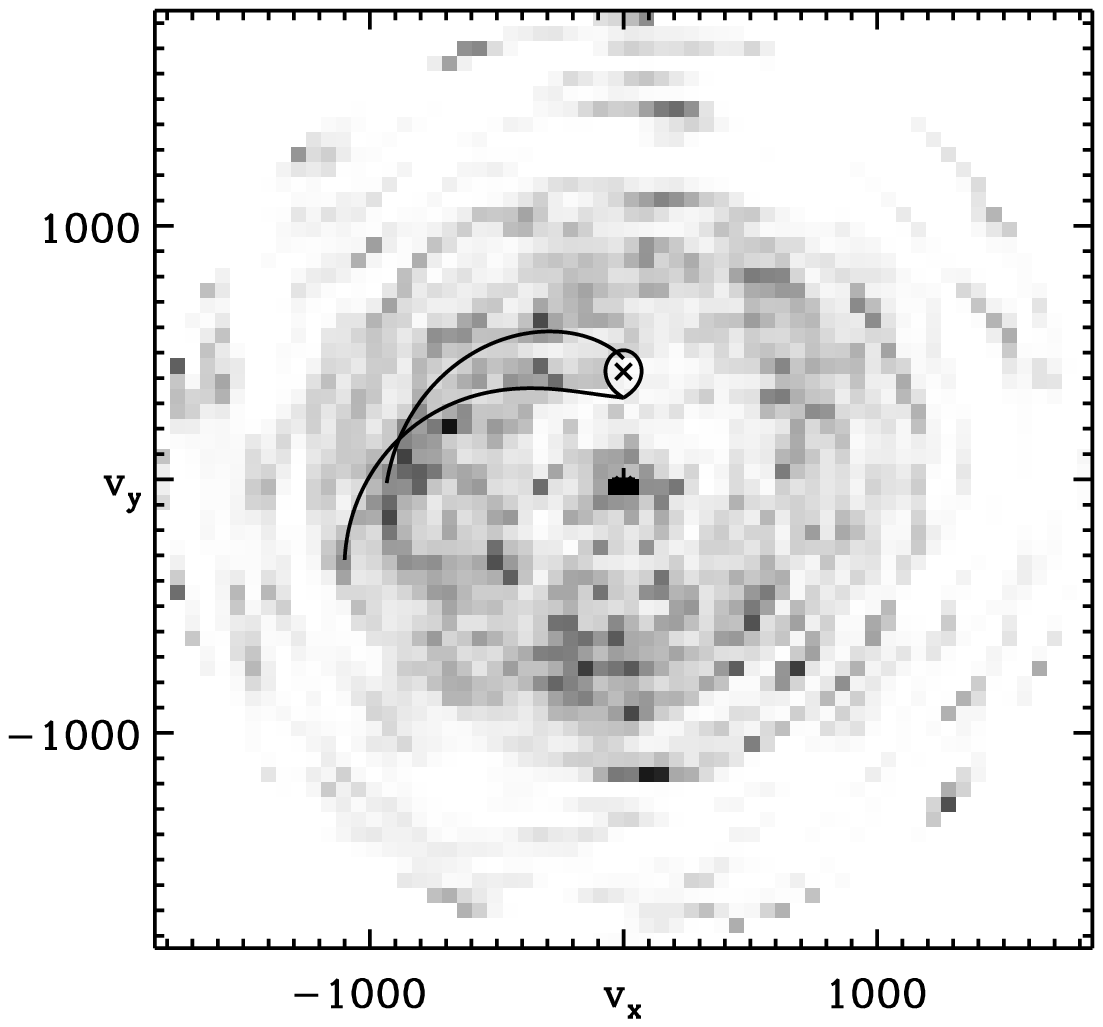}
\caption{Reconstructed projected velocity maps (in km~s$^{-1}$) for
the He{\sc ii} (left) and H$\alpha$ (right) emission during the April
20 observation (see Fig.~\ref{fig:prof}). Also shown are the location
of the binary components (crosses), the trajectory of the accretion
stream from the secondary and corresponding Keplerian velocity for
$q=0.05$ and $K_2=420$~km~s$^{-1}$. For this particular choice of
parameters, and for an appropriate orbital reference phase, the
emission region in the lower-left quadrant is associated with
stream-disk interaction (the hotspot).\label{fig:map}}
\end{figure}

Despite the incomplete phase coverage and the absence of a reference
for superior conjunction of the secondary, we attempted to locate the
emission site of the S-wave in the binary velocity plane using the
Doppler tomography technique (Marsh \& Horne 1988).  Maps were
reconstructed for He{\sc ii} and H$\alpha$ for the April 20
observations using the software developed by Spruit (1998), assuming
an orbital period of 0.1708~day and a null systemic velocity. It was
not possible to combine this dataset with our other observations as
this would have required $P_{\rm orb}$ to be known much more
accurately ($\Delta P_{\rm orb}/P_{\rm orb}= \Delta \phi P_{\rm orb}/
T \sim 10^{-5}$ day with $T\approx90$ days and an error on the phase
set by the exposure time of $\Delta \phi=0.01$). { Furthermore, the
0.1708~day modulation is probably related to superhumps so that the
orbital period can be expected to be close to this value with
an uncertainty of a few percent. We tried several different orbital
periods around 0.17~day but this had no significant impact on the
maps. } The reconstructed trailed profiles and corresponding velocity
maps are shown in Fig.~\ref{fig:prof} and Fig.~\ref{fig:map}
respectively.

Each pixel on the projected Doppler map corresponds to an S-wave in
the trailed spectra. The binary components are located on the y-axis
of the map at positions corresponding to their semi-amplitude
velocities $(v_x,v_y)=(0,-K_1)$ for the primary and $(0,K_2)$ for the
companion (indicated by crosses). However, since the orbital reference
phase is arbitrary (here we chose $t_0=$51654.420 in MJD), the map can
be rotated in any fashion around the velocity origin.  The secondary
position, the ballistic gas stream trajectory and the corresponding
Keplerian velocity at the location of the stream shown on
Fig.~\ref{fig:map} result from particular choices of $K_2$ and
$q=M_2/M_1=K_1/K_2$ discussed further below.

The trailed emission of He{\sc ii} shows a nice S-wave pattern which is
consistent with the reported 0.1708~day photometric period (Cook et al.
2000; Patterson 2000).  Most of the He{\sc ii} emission originates from a
localized region but there is also an underlying (double-peaked) disk
component which is visible in the original data. On the other hand, the
disk dominates the H$\alpha$ profile producing a ring of emission in the
velocity map. We also note the regions of enhanced H$\alpha$ emission at
about (-910,0)~km~s$^{-1}$ (consistent with the He{\sc ii} S-wave) and
$(-210,-730)$~km~s$^{-1}$. The peak at the velocity origin (the center of
mass) is due to the H$\alpha$ emission component at rest wavelength
discussed in \S3.2.

The bright emission region is typically associated with emission
arising from the stream-disk interaction region or from the X-ray
heated surface of the secondary (e.g. Smak 1985; Marsh, Robinson \&
Wood 1994; Casares et al. 1995ab; Harlaftis et al. 1996, 1997ab; Hynes
et al. 2000a). Some SXTs also have low-velocity emission regions in
their doppler maps, possibly associated with a magnetic propeller
(Hynes et al. 2000a). Here, the velocity of $\sim 900$~km~s$^{-1}$
makes it doubtful that we observe such a phenomenon.

Assuming the He{\sc ii} emission arises from the heated hemisphere of
the secondary, the semi-amplitude $900$~km~s$^{-1}$ {
(underestimating the real $K_2$)} gives a rough constraint on the mass
function through :
\begin{equation}
f(M)=\frac{M_1^3\sin^3 i}{(M_1+M_2)^2}=\frac{K_2^3 P_{\rm orb}}{2 \pi G}
\label{eq:fm}
\end{equation}
This gives a {\em minimum} mass for the primary of 12 M$_\odot$, which
{ would make XTE~J1118+480 at least as massive as the black hole in
V404 Cyg. Lower inclinations would imply an even greater mass making
this system rather unusual compared to the dynamical masses inferred
in other SXTs (see, e.g., van Paradijs \& McClintock 1995).  Although
we cannot formally reject this possibility, the He{\sc ii} emission is
unlikely to originate from the heated surface of the secondary.}

\begin{figure}
\plottwo{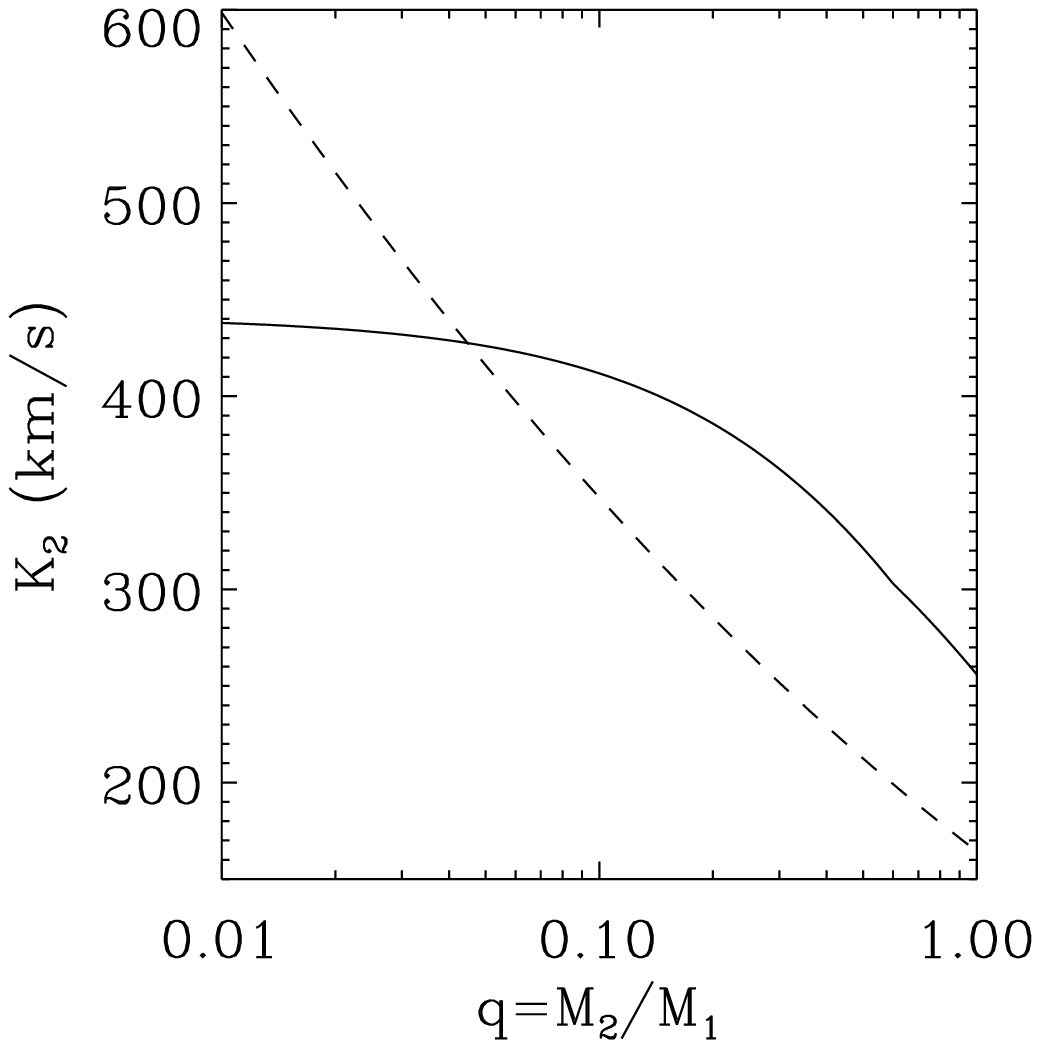}{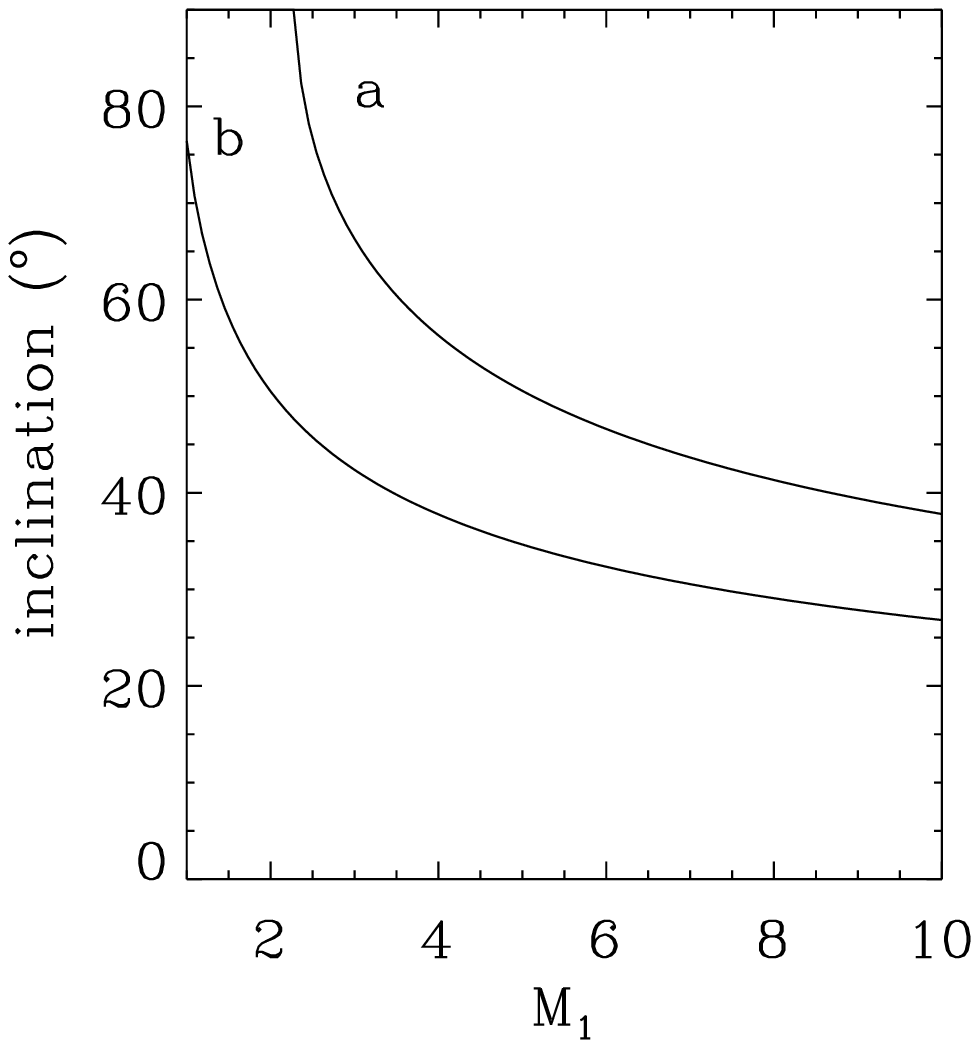}
\caption{Left: Estimates of $K_2(q)$. The solid line was obtained by
assuming that the peak-to-peak velocity of the H$\alpha$ line
represents the velocity of the outer radius $R_{\rm d}$ of a Keplerian
disk, with $R_{\rm d}(q)$ given by theoretical arguments on disk
truncation. The dashed line was obtained by forcing the ballistic gas
stream trajectory and its corresponding Keplerian velocity to
intersect at the location of the He{\sc ii} emission region in the
Doppler map (Fig.~\ref{fig:map}).  Given the (large) uncertainties
involved, we consider the agreement acceptable for $0.02\lsim q \lsim
0.1$.  Right: Estimate of the inclination $i$ as a function of the
primary mass $M_1$. Given $K_2$ and $q$, $i(M_1)$ is obtained from the
mass function (Eq.~\ref{eq:fm}). The two curves correspond to the
extreme values of $q$ and $K_2$ derived from the left panel ($a$ for
$q=0.02$ and $K_2=500$~km~s$^{-1}$; $b$ for $q=0.1$ and
$K_2=350$~km~s$^{-1}$). Values in between the two curves are
acceptable within our assumptions. The high inclinations required for
small values of $M_1$ (in solar masses) are necessary to explain the
high peak-to-peak velocity (hence outer disk velocity) in the data.
\label{fig:cons1}}
\end{figure}

The He{\sc ii} emission is more likely to originate from the
stream-disk interaction. { Typically, the hotspot is located (see
above references) somewhere between the ballistic gas stream
trajectory and its corresponding Keplerian velocity (which defines a
circularization radius)}. We derive a { lower limit on $K_2(q)$
(hence a lower limit on the primary mass)} by assuming that the
emission is located at the intersection of the two trajectories. The
values of $K_2$ and $q$ for which the intersection matches the He{\sc
ii} emission are plotted as a dashed line in the left panel of
Fig.~\ref{fig:cons1}.

A comparison between the two estimates of $K_2(q)$ shows that values
of $0.02\lsim q\lsim 0.1$ best fit the data in the sense that they
minimize the discrepancies. Formal agreement requires $q\approx0.045$
($K_2\approx430$~km~s$^{-1}$) but this is, of course,
model-dependent. Knowing $q$ and $K_2$, the mass function
(Eq.~\ref{eq:fm}) gives the inclination $i$ as a function of $M_1$.
This relation is plotted in Fig.~\ref{fig:cons1} (right panel) for the
two extreme cases ($q=0.02$,$K_2=500$~km~s$^{-1}$) and
($q=0.1$,$K_2=350$~km~s$^{-1}$). Clearly, a NS would require large
inclinations while a BH gives intermediate inclinations.  { Given
the assumptions made for the interpretation of the peak-to-peak
velocity and the He{\sc ii} emission, we are likely to underestimate
$K_2$ in both cases. This would only strengthen the case for a black
hole primary.}

The stream-disk interaction apparently extends some way along the
Keplerian velocity trace as previously seen in other SXTs
(e.g. Harlaftis et al. 1996). This could also be a numerical artifact
{ possibly due to an inaccurate orbital period}. The second weaker
emission region in the H$\alpha$ velocity map could be due to
continued disturbance of the accretion disk further along in azimuth.
In this picture the stream-disk interaction takes place inside the
disk as emphasized by the larger velocity amplitude of the S-wave with
respect to the peak-to-peak amplitude of the Balmer emission
lines. This interpretation suggests significant stream overflow above
the outer disk and an interaction region close to the circularization
radius. { Alternatively, systematic effects leading to lower
observed peak-to-peak velocities (e.g. local line broadening, Marsh
1998) would place the hot spot closer to the disk outer edge}.

\subsection{Interstellar Absorption}
\begin{figure}
\plottwo{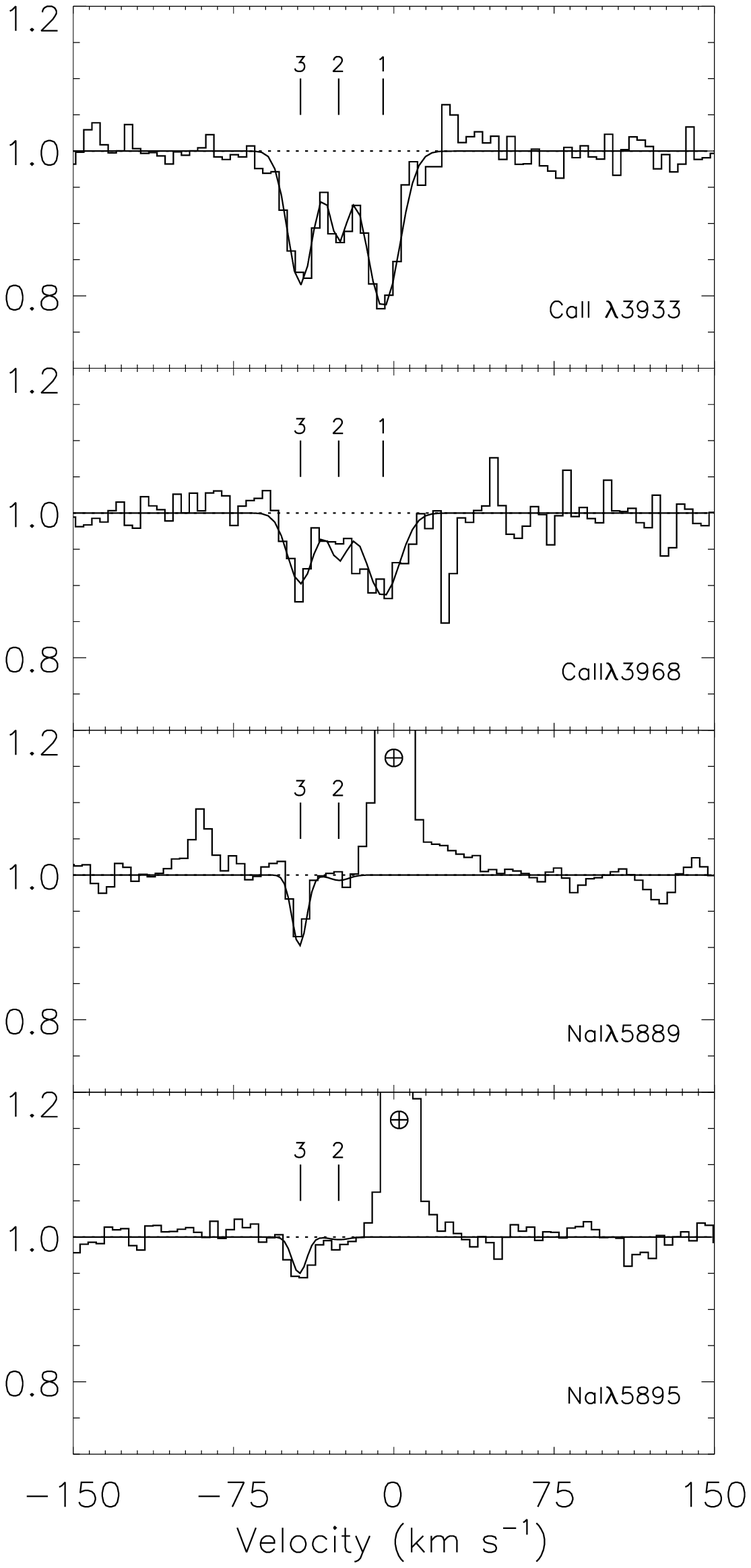}{blank.eps}
\caption{Echelle spectroscopy of XTE~J1118+480 reveals three narrow
interstellar Ca{\sc ii}~$\lambda 3933.6$ absorption lines (seen also
in Ca{\sc ii}~$\lambda 3968$ and Na{\sc i}~$\lambda\lambda
5889,5895$). The total equivalent widths are $0.108 \pm 0.009 \AA$ for
the Ca{\sc ii}~$\lambda 3933$ lines and $0.061 \pm 0.008 \AA$ for the
Ca{\sc ii}~$\lambda 3968$ lines. All the absorption features shown in
the four panels above (summed Echelle spectra) are fully consistent
with each other, in width and velocity (note the Na{\sc i} sky
emission at rest wavelength).  In each panel, the data are shown as an
histogram, the solid line shows a fit to the absorption lines and the
dotted line a fit to the continuum. The three Ca{\sc ii} lines can be
used to constrain the low hydrogen absorption column and distance to
the source (see text for details).
\label{fig:ca2}}
\end{figure}

The Echelle spectra of XTE~J1118+480 obtained on April 7, 2000 reveal
the presence of weak Ca{\sc ii}~$\lambda 3933$ absorption features,
which can be used to estimate the neutral hydrogen absorption column
to the source. The spectra also show Ca{\sc ii}~$\lambda 3968$ in
absorption, with an equivalent width of order half that of Ca{\sc
ii}~$\lambda 3933$ features, as expected for this doublet (see, e.g.,
Cohen 1975). Both features are seen on two overlapping orders with
similar structure. The Ca{\sc ii} lines are detected in some of the
individual Echelle exposures with identical profiles.
Figure~\ref{fig:ca2} shows the Ca{\sc ii}~$\lambda\lambda 3933,3968$
absorption features in the summed Echelle spectra. The two orders have
also been summed for this plot with weighting corresponding to the
different sensitivities at the locations of the Ca{\sc ii} lines.

A Gaussian fit to each of the three absorption components yields
heliocentric velocities $v_0 \simeq -44$, $-26$ and $-5$~km~s$^{-1}$.
These velocities are consistent with an absorption component clearly
seen at $\sim -44$~km~s$^{-1}$ and the hint of another component at
$-26$~km~s$^{-1}$ in NaI~$\lambda\lambda 5889, 5895$. The presence
of strong Na emission at the rest wavelength (sky emission) precludes
the detection of the presumed third absorption component at
$-5$~km~s$^{-1}$ and the use of Na absorption for the determination of
the H{\sc i} absorption column, N$_{\rm H{\sc i}}$, to XTE~J1118+480.

We proceed as follows to determine N$_{\rm H{\sc i}}$.\footnote{Note
that we do not have to worry about a stellar Ca{\sc ii} absorption
component from the companion star because the emission of
XTE~J1118+480 in outburst is presumably dominated by accretion} We
assume that the gas in each absorbing cloud has a Maxwellian velocity
distribution centered on its heliocentric velocity $v_0$, so that the
Ca{\sc ii} optical depth of the cloud can be described by (Spitzer
1978):
\begin{equation}
\tau [v-v_0] = {\rm N_{Ca{\sc ii}}} \frac{\sqrt{\pi} e^2 f \lambda_0}{m_e c b}
e^{-[(v-v_0)/b]^2},
\end{equation}
where ${\rm N_{Ca{\sc ii}}}$ is the total cloud absorption column,
$\lambda_0 = 3933.6~\AA$ is the rest wavelength of the Ca{\sc ii}~K
line, $f=0.688$ is the oscillator strength of this transition
(Cardelli \& Wallerstein 1986), $b$ is the velocity dispersion
of the cloud, $e$ is the charge of the electron, $m_e$ is the mass of
the electron and $c$ is the speed of light. Relative to a unity
continuum, the line profile of a given absorbing cloud component has a
depth given by:
\begin{equation}
e^{- \tau[v-v_0]}
\end{equation}
at any speed $v$ (or equivalently wavelength) around the central value
$v_0$.

To deduce precise values of $b$ and ${\rm N_{Ca{\sc ii}}}$, we have
compared the observed Ca${\sc ii}$ line profiles with theoretical,
multi-component absorption line profiles. Initial profiles were
calculated using the values of $b$ and ${\rm N_{Ca{\sc ii}}}$ derived
from Eqs.~(6) \&~(7), which were then convolved with a Gaussian
instrumental Line Spread Function whose FHWM was measured from the arc
lines used to wavelength calibrate the data. The best values of $b$
and ${\rm N_{Ca{\sc ii}}}$ were found by minimizing $\chi^2$ between
the data and the theoretical fits. The resulting values for the
Ca${\sc ii}$ lines are: $b\:=\:9.1, 2.4 $ and 6.2~km~s$^{-1}$, $\log
{\rm N_{Ca{\sc ii}}} = 11.85, 11.34$ and 11.67 for the components at
$v_0\:=\:-5, -26$ and $-44$~km~s$^{-1}$ respectively.

The translation of the Ca{\sc ii} absorption column to an hydrogen
absorption column is somewhat delicate, especially in view of the
possibly efficient deposition of gaseous Ca{\sc ii} onto dust grains
(see, e.g., Vallerga et al. 1993). However, there is a substantial
body of observational data on Ca{\sc ii} absorption which allows us to
use an average value for the ratio ${\rm N_{Ca{\sc ii}}/N_{H{\sc
i}}}$. In their independent studies, both Cohen (1975) and Sembach \&
Danks (1994) find an average average value of $\log[{\rm N_{Ca{\sc
ii}}/N_{H{\sc i}}}] \simeq -8.3$ for high latitude line-of-sights,
with approximate errorbars of $\pm 0.2$ in log in the two
cases. Assuming this average is representative of the line of sight to
XTE~J1118+480, this translates in a value $\log[{\rm N_{H{\sc
i}}~(cm^{-2})}] = 20.45 \pm 0.2$, where the errorbars come from the
conversion of ${\rm Ca{\sc ii}}$ to ${\rm H{\sc i}}$ absorption.

{
We note that this estimate is somewhat larger than the values
$\log[{\rm N_{H{\sc i}}~(cm^{-2})}] \approx 20$ considered as
reasonable by Hynes et al. (2000; see also McClintock et al. 2000;
Esin et al. 2000) when dereddening the low-energy spectrum of
XTE~J1118+480. This discrepancy raises the possibility that the value
obtained from the Ca${\sc ii}$ measurement overestimates the true
value of N$_{H{\sc i}}$ on the line of sight to the source. Most
likely, this would come from the conversion from Ca${\sc ii}$ to
H${\sc i}$ which may not be ``average'' as assumed above for
XTE~J1118+480. It is also possible to measure the total $N$(H~I) along
the sightline towards XTE~J1118+480 from 21~cm emission
observations. Using data from the Leiden/Dwingeloo Survey (Hartmann \&
Burton 1997), the H~I column at ($l,b$)=(157.5,62.0) is
$1.32\times10^{20}$~cm$^{-2}$, a factor of two lower than we
derive. This value represents an {\it upper} limit of course, since
the 21~cm emission is integrated along the line of sight through the
entire Milky Way. This therefore also suggests our value of $N$(H~I)
is over-estimated, although the Dwingeloo beam size of 36$'$ is too
large to measure higher $N$(H~I) over smaller angular scales, and it
remains possible that a small, dense knot of H~I with $N$(H~I)$>>
1.3\times10^{20}$~cm$^{-2}$ could be lying directly along the line of
sight. }

It is encouraging to see that the properties derived above for the
clouds are consistent with the average properties of high latitude
line-of-sights described in Sembach \& Danks (1994): $\langle{\rm
N_{Ca{\sc ii}}}\rangle =4.3 \pm 1.3 \times
10^{11}$~cm$^{-2}$. Furthermore, these authors find that, along
extended sight lines at high latitude, the average number of absorbing
Ca{\sc ii} clouds is 3.6~kpc$^{-1}$. This indicates that XTE~J1118+480
is located at a distance $\approx 0.83$~kpc if the line of sight to
the source has average properties, an estimate which is consistent
with those for other SXTs. The actual distance to XTE~J1118+480 could
be somewhat different if this line-of-sight has properties deviating
significantly from the mean.  Note that our estimate of the distance
is in rough agreement with that of Uemura et al. (2000) who use the
18.8 quiescent magnitude to derive $d\sim0.5$~kpc (M-type secondary)
or $d\sim1.5$~kpc (K secondary), and is consistent with the 0.8~kpc
distance estimate of McClintock et al. (2000), assuming that the
primary is a massive BH and that a third of the light in quiescence is
provided by the M-dwarf secondary. { It is also of interest to note
that the 21~cm emission from the Leiden/Dwingeloo Survey mentioned
above shows two peaks in brightness temperature which correspond
closely to the Ca~II components at $-5$ and
$-44$~km~s$^{-1}$. Although the distance of the H~I is not known, the
scale height of H~I in the Milky Way is $\sim 1$~kpc, and since the
21~cm emission arises from {\it all} the H~I in the Galaxy along the
sightline, XTE~J1118+480 must be sufficiently far away so as to show
the Ca~II absorption from both H~I components. This again suggests a
significant distance between us and XTE~J1118+480.}

\section{Discussion}

\subsection{The absorption troughs}

An interesting feature of the spectra are the absorption troughs in
$H\gamma$ and $H\beta$. Although rarely discussed (see, however, Soria
et al. 2000), these have been observed in several other transients
with similar FWHM, { including} GRO J1655-40, A0620-00, GS1124-68,
Nova Mus 91, XTE J2123-058, XTE J1859+226, and especially GRO J0422+32
which bears many similarities to XTE J1118+480 (see conclusion).

This type of line profile is also seen in dwarf novae (DN) and {
nova-like} (NL), systems which are analogous to low-mass X-ray
binaries (LMXBs), except for a primary which is a white dwarf.
Typically, the spectrum of quiescent DN display strong emission lines
which are gradually replaced by absorption features during the rise to
outburst and vice-versa during the decay (e.g. Szkody et al. 1990).

Schematically, the absorption is thought to arise from the
optically-thick accretion disk whilst the emission is thought to
result from photoionization in a chromosphere-like optically thin
region above the disk. A key feature is that the expected spectrum is
inclination dependent since the absorption should disappear as the
system gets closer to being edge-on due to strong line limb-darkening
in the 2D disk (see e.g. Marsh \& Horne 1990; la Dous 1989; Wade \&
Hubeny 1998). There is indeed such an observed dependence in both DN
and NL with the highest inclination systems showing only emission
lines (la Dous 1991).

Although the detected absorption makes it unlikely that the
inclination of XTE~J1118+4800 is very high, this constraint is
weak. For instance, GRO J1655-40 also showed Balmer absorption (Soria
et al. 2000) but has a well-determined inclination of 70\arcdeg. In
contrast with DN and NL, irradiation heating is much stronger in SXT
where it usually dominates the heat balance in the outer disk regions
(hence the optical emission, van Paradijs \& McClintock 1994). Energy
deposited by soft X-rays easily causes a thermal inversion in the top
layers where the emission lines originate (e.g. Ko \& Kallman
1994). If the optically thin atmospheres of SXTs are indeed more
extended than in DN, one would expect Balmer absorption not to be as
common an occurence as in the non-magnetic cataclysmic variables.

The spectral models of X-ray irradiated disks computed by Sakhibullin
et al. (1998) show that softer irradiation leads to the disappearance
of the absorption troughs. As the irradiation spectrum hardens, the
X-ray photons deposit their energy in deeper disk layers, which do not
contribute to the line emission. Inversely, any parameter change
resulting in additional heating of the disk atmosphere leads to more
emission. Any combination of a low inclination, low X-ray luminosity,
hard X-ray spectrum or low fraction of reprocessed X-ray photons in
the disk could explain the absorption troughs.

Those SXTs in which Balmer absorption has been observed have probably
been guilty of one or several of the above.  Interestingly, in GRO
J0422+32, the H$\alpha$ and H$\beta$ lines evolved from absorption to
emission within 3 days during the rise to one of the mini-outbursts
(Callanan et al. 1995). Although the authors could not find any
evidence for increased X-ray flux between 0.5-10 keV, the simultaneous
rise in He{\sc ii} emission does suggest a larger flux of soft
0.05-0.3~keV photons which would have a strong effect on the
chromosphere. However, a hard X-ray upturn in GRO J1655-40 was
accompanied by a dramatic increase of H$\alpha$ emission but also by a
{\em decrease} of the He{\sc ii} $\lambda 4686$ flux (Shrader et
al. 1996). It seems difficult to identify the dominant culprit but we
propose the low X-ray flux (see next section) and the hard X-ray
spectrum of XTE~J1118+480 as prime suspects responsible for the
absorption troughs observed.

\subsection{The optical magnitude}

A unique characteristic of this system is the large optical to X-ray
ratio, as compared to other SXTs. The ratio for XTE~J1118+480 is not
dissimilar (Garcia et al. 2000) to that found in Accretion Disk Corona
(ADC) sources where the X-ray flux is partly hidden by the disk seen
edge-on.  Yet, XTE~J1118+4800 has not shown any of the typical
behaviour associated with ADC sources, namely X-ray dips or
eclipses. In addition, our crude `fitting' of the peak-to-peak
velocity of the Balmer lines and of the velocity map rather hints at
intermediate inclinations.

This high ratio might simply be due to the low absorption and
proximity of the source. At the $\sim$ 0.8~kpc distance implied by the
Ca{\sc ii} absorption, the 2-10~keV X-ray flux ($\approx
8\cdot10^{-10}$erg~s$^{-1}$~cm$^{-2}$, Yamaoka et al. 2000) is such
that $L_{\sc x}\approx 6\cdot 10^{34}$~erg~s$^{-1}$.  By analogy with
the low/hard state of Cyg X-1, Fender et al. (2000) find the broad
band X-ray luminosity is $\sim 10^{36}$erg~s$^{-1}$~cm$^{-2}$
at 1~kpc, i.e. this outburst was weak for a SXT.

We compare the X-ray flux and optical magnitude $M_V$ by computing the
quantity $\Sigma=(L_{\sc x}/L_{\rm Edd})^{1/2}(P_{\rm orb}/1{\rm
~hr})^{(2/3)}$, where $L_{\rm Edd}= 1.3 \times 10^{38}
(M_1/M_\odot)$~erg~s$^{-1}$ is the Eddington luminosity.  With $P_{\rm
orb}=0.1708$~day, we find $\log\Sigma\lsim -1.2$ with
$M_1=1$~M$_\odot$ while $M_V\approx 3.5$ for $V=13$ and
$E(B-V)=0$. Within the error bars, this is consistent with the
correlation found for other SXTs between $\log\Sigma$ and $M_V$ by van
Paradijs \& McClintock (1994). In other words, the optical flux of
XTE~J1118+480 is roughly in agreement with what would be expected if
the disk is irradiation-dominated.

The agreement is less good for higher $M_1$, the optical flux being
larger than expected from the correlation. The X-ray heated companion
could contribute to the optical as in, e.g., the persistent LMXB Cyg
X-2. { In this case we might expect the He{\sc ii} emission to
originate from the X-ray heated surface of the secondary. We have
argued this is unlikely. A better candidate for this extra optical
emission is} synchrotron emission (Hynes et al. 2000b) originating in
a magnetic corona (Merloni et al. 2000),\footnote{Note that the corona
model of Merloni et al. 2000 may be challenged by the observations of
Haswell et al. (2000c) which show that the near-UV variability lags
behind the X-ray variability by 1 to 2 seconds.} an
advection-dominated accretion flow (ADAF; Esin et al. 2000) and/or a
jet ($\sim$ equivalent to the first model with the base of the jet
acting as the magnetic corona; { Markoff et
al. 2000}). XTE~J1118+480 { probably has} a powerful compact jet,
with emission extending at least to the near-infrared (Fender et
al. 2000). Both ADAF and magnetic corona models can explain the high
optical/X-ray ratio but the EUV predictions may differ. The model
presented by Merloni et al. (2000) has a large blackbody component
peaking at $\nu\approx10^{16.5}$~Hz due to hard X-ray reprocessing in
the disk.  { The soft X-ray observations of {\it Chandra} do not
seem to support this model (McClintock et al. 2000; Esin et
al. 2000).}

\subsection{Superhumps ?}

The detection of superhumps (Uemura 2000) could add several
constraints to the system parameters.  According to current models,
superhumps appear when the disk can reach the 3:1 resonance radius,
requiring small mass ratio $q\lsim0.3$ (Warner 1995). In principle,
the relative difference between the superhump period and the orbital
period can give an even better constraint on $q$ (e.g. O'Donoghue \&
Charles 1996). Such a small mass ratio is compatible with the
constraint $0.02\lsim q\lsim 0.1$ that we derived in \S3.5.

In DN and NL, the origin of the optical variation could lie in the
enhanced viscous dissipation modulated on the superhump period.
However, these variations should be swamped in SXTs where irradiation
heating rather than viscous heating dominates the optical
output. Haswell et al. (2000a) recently argued that the small change
in the accretion disk area on the superhump cycle, rather than the
enhanced viscous dissipation, would give the desired effect for
SXTs. The amplitude of the superhump would however decrease with
larger inclinations.

The detection of superhumps in XTE~J1118+4800, if confirmed, would
therefore either mean the disk was not irradiation-dominated (which is
unlikely considering, e.g., the correlated X-ray/near-UV behaviour
reported by Haswell et al. 2000c) or that the inclination is low
enough for the variations of the area of the disk to be observable.

\subsection{Is XTE~J1118+480 a halo object ?}

The high Galactic latitude of XTE~J1118+480 ($b=+62^o$) is in sharp
contrast with the latitudes of other LMXBs which cluster next to the
Galactic plane. Whether or not XTE~J1118+480 is a halo object
obviously depends on its actual distance.  At our inferred $d\sim
0.8$~kpc, XTE~J1118+480 lies $\sim 0.7$~kpc above the Galactic plane.

The distribution of BH LMXBs around the Galactic plane has a
dispersion of $\sim 0.5$~kpc. The larger dispersion of NS LMXBs ($\sim
1$~kpc) is probably due to larger kick velocities at birth (White \&
van Paradijs 1996). Whether XTE~J1118+480 has a BH or a NS primary,
its height above the Galactic plane is not much larger than for other
LMXBs. For instance, White \& van Paradijs (1996) find $z\approx
0.9$~kpc for the BH Nova Oph 1977.

A low metallicity environment would be supported by the low Bowen to
He{\sc ii} flux ratio $\approx 0.3$ (Table.~\ref{tab:lines}) that we
find (Motch \& Pakull 1989). But the detection of N and the
non-detection of C and O lines in the UV spectra has been used to
argue that the matter has been CNO processed in the companion (Haswell
et al. 2000b) and that, from an evolutionary point of view,
$M_1<3$~M$_\odot$ is preferred to allow for an evolved
secondary. Considering that XTE~J1118+480 is not extremely far from
the distribution of other LMXBs above the Galactic plane and that CNO
processing might explain the low Bowen/He{\sc ii} flux ratio, we
conclude that there is no compelling evidence for a halo origin of
XTE~J1118+480.

\section{Conclusion}

Let us first summarize the main conclusions of this work.

$\bullet$ The optical spectrum of XTE~J1118+480 in outburst shows
several variable broad and weak emission lines superposed on a
$F_\nu\propto \nu^{1/3}$ continuum typical of an optically-thick
accretion disk.  The He{\sc ii} line shows a strong S-wave pattern
which is consistent with the claimed 0.1708~day photometric period.
We see no { other} obvious periodic or long-term behaviour in the lines.

$\bullet$ We find from the Ca{\sc ii} lines that the interstellar
absorption to the source is low ($\log[{\rm N_{H{\sc i}}~(cm^{-2})}] =
20.45 \pm 0.2$) and that the presence of three intervening clouds
suggests a distance of $\sim 0.8$~kpc to the source. Since the low
Bowen to He{\sc ii} ratio could indicate CNO-processed material
(Haswell et al. 2000b) rather than an intrinsically low metallicity,
we conclude there is no strong evidence to support a halo origin for
XTE~J1118+480.

$\bullet$ { We estimate an upper limit for the 0.05-0.3~keV source
flux of $\approx 10^{-8}$~erg~s$^{-1}$~cm$^{-2}$. The dereddened EUVE
spectrum (using a column absorption $\log[{\rm N_{H{\sc
i}}~(cm^{-2})}] \approx 20$; Hynes et al. 2000; Mcclintock et
al. 2000) appears compatible with this value.  This suggests that a
significant fraction of the source flux is emitted in the EUV/Soft
X-ray band. Given the very strong dependence of the slope of the
dereddened EUVE spectrum on ${\rm N_{H_{\sc i}}}$ , it may be
difficult to accomodate a higher value for ${\rm N_{H_{\sc i}}}$ as
derived from Ca{\sc ii} absorption in \S3.6.

}
$\bullet$ The absorption in H$\gamma$, H$\beta$ and the Balmer jump
(Hynes et al. 2000b), originating from the optically-thick disk,
suggests that X-ray heating of the atmosphere is not as strong as in
other SXTs, which may be consistent with the (very) low X-ray flux and
hard X-ray spectrum. This also disfavors very high inclinations for
which limb-darkening would remove the absorption features.

$\bullet$ The Balmer lines have peak-to-peak velocities of $\approx
1250$~km~s$^{-1}$. Assuming this corresponds to the Keplerian velocity
of the outer disk, we use theoretical arguments on the disk size to
estimate $K_2(q)$. We find $250 \lsim K_2$~(km~s$^{-1}$)$\lsim 450$
for $0.01<q<1$.

$\bullet$ Using the reconstructed Doppler tomogram, we tentatively
identify the He{\sc ii} S-wave with the stream-disk interaction. This
gives us a second estimate of $K_2(q)$ which, combined with the first
one, restricts possible values for $q$ to $0.02\lsim q\lsim0.1$.  This
further implies that if the primary is a NS, the system should have a
high inclination ($i\gsim70$\arcdeg), while for a BH primary, an
intermediate inclination is preferred (30\arcdeg$\lsim i
\lsim$50\arcdeg\ for $10\gsim M_1/M_\odot\gsim4$).

Superhumps (Uemura 2000), if they require $q<0.3$ and low to
intermediate inclinations (Haswell et al. 2000a), the absence of dips
or eclipses in the X-ray lightcurve, the lack of high frequency QPOs
(Revnivtsev et al. 2000) and the broad Ly$\alpha$ line (Haswell et
al. 2000b) all favor an intermediate inclination system containing a
BH.  All of the arguments above are, however, rather uncertain and
model-dependent. { The assumptions made to interpret the
peak-to-peak velocity and He{\sc ii} emission probably underestimate
$K_2$. A larger $K_2$ would strengthen the case for a black hole
primary}.

We conclude by emphasizing the similarities between XTE~J1118+480 and
the black hole LMXB GRO~J0422+32 in mini-outburst: both have orbital
periods of $\sim 4$ hours, stayed in the low/hard X-ray state during
outburst, had low X-ray luminosities, showed superhumps and had
identical optical spectra (including the absorption troughs). Scaling
the $V=19.5$ secondary of GRO~J0422+32 from $d\approx2.5$~kpc to the
distance of XTE~J1118+480 ($\approx 0.8$~kpc) gives $V=18.5$,
consistent with the reported 18.8 quiescent magnitude. The system
parameters for GRO~J0422+32 (taken from Chen et al. 1997) are also
within our allowed range for a coherent explanation of both the Balmer
line peak-to-peak velocity and the location of the He{\sc ii} emission
in XTE~J1118+480. Quiescent studies, which can constrain the spectral
type of the secondary and $K_2$, will shed more light on this
interesting object.

\section*{Acknowledgments}

The authors are grateful to Mike Garcia and Jeff McClintock for very
useful discussions, to Todd Tripp for sharing his knowledge of the
APO Echelle spectrograph and to { Rob Hynes for a very useful
referee report}. We are indebted to Ed Turner for granting us
some of his Director's discretionary time at APO.  Support for this work
was provided by NASA through Chandra Postdoctoral Fellowship grant
number PF9-10006 awarded by the Chandra X-ray Center, which is
operated by the Smithsonian Astrophysical Observatory for NASA under
contract NAS8-39073. GD acknowledges support from the European
Commission through the TMR network `Accretion on to Black Holes,
Compact Stars and Protostars' (contract number ERBFMRX-CT98-0195) and
from the Leids Kerkhoven-Bosscha Fonds. RSJK acknowledges support from
NSF grant AST96-16901 and the Princeton University Research Board. PS
acknowledges support from NASA grant NAG 5-7278.


\clearpage

\begin{table*}
\caption{ SPECTROSCOPIC OBSERVATIONS OF XTE~J1118+480}
\begin{center}
\begin{tabular}{ccccc} \hline \hline
\\
Date (UT) & Epoch & Instrument & Wavelength coverage & Exposures \\
& (HJD-2400000) & & (\AA) &\\
\\
\hline
\\
April 7, 2000 &  51641.7 & DIS lowres & 4000-9000& $2  \times 60$s\\
 & & Echelle & 3500-9800 & $2 \times 600$s + $12 \times 1200$s\\
April 20, 2000 & 51654.9 & DIS hires & 4200-5000 / 5800-6800 & $26 \times 300$s \\
April 23, 2000 & 51657.8 & DIS hires & 4200-5000 / 6200-7200 & $1 \times 300$s \\
May 6, 2000 & 51670.8 & DIS hires & 4200-5000 / 6300-7300 & $1 \times 480$s \\
May 12, 2000 & 51676.8 & DIS hires & 4200-5000 / 6300-7300 & $1 \times 300$s \\
May 15, 2000 & 51679.7 & DIS hires & 4200-5000 / 6300-7300 & $1 \times 300$s \\
May 28, 2000 & 51692.8 & DIS hires & 4200-5000 / 6300-7300 & $1 \times 300$s \\
July 3, 2000 & 51728.7 & DIS hires & 4200-5000 / 6300-7300 & $1 \times 300$s \\
July 4, 2000 & 51729.7 & DIS hires & 4350-5150 / 6050-7050 & $9 \times 300$s + $ 2 \times 600$s \\
\\
\hline
\end{tabular}
\label{tab:obs}
\end{center}
\end{table*}

\begin{table*}
\caption{EMISSION LINES AND ABSORPTION TROUGHS IN XTE~J1118+480}
\begin{center}
\begin{tabular}{lcccc} \hline \hline
\\
Line & $\lambda_0$ (\AA) & $\lambda_{\rm obs}-\lambda_0$ (\AA) & EW (\AA) & FWHM (1000 km~s$^{-1}$)\\
\\
\hline
\\
H$\gamma$       & 4340.5 & -2$\pm$4  & -1.1$\pm$0.4 & 2.1$\pm$0.5\\
	        & 4340.5 & 15$\pm$5  &  1.8$\pm$0.5 & 3.5$\pm$0.5\\
He{\sc i}	& 4471.5 & 1$\pm$5   & -0.3$\pm$0.2 & 1.9$\pm$0.6\\
		& 4471.5 & 4$\pm$5   &  0.7$\pm$0.4 & 4.0$\pm$0.8\\
Bowen blend 	& 4638   & 8$\pm$10  & -0.5$\pm$0.3 & 2.0$\pm$1.1\\
He{\sc ii}     	& 4686.7 & 0$\pm$3   & -1.5$\pm$0.4 & 2.3$\pm$0.4\\
H$\beta$	& 4861.3 & -1$\pm$5  & -1.0$\pm$0.3 & 1.8$\pm$0.3\\
		& 4861.3 & 13$\pm$6  &  1.5$\pm$0.5 & 3.1$\pm$0.4\\
He{\sc i} 	& 4921.9 & 2$\pm$5   & -0.2$\pm$0.2 & 1.4$\pm$0.6\\
H$\alpha$	& 6562.8 & 0$\pm$2   & -1.9$\pm$0.4 & 1.8$\pm$0.4\\
He{\sc i} 	& 6678.1 & -1$\pm$5  & -0.6$\pm$0.4 & 1.8$\pm$0.8\\
\hline
\end{tabular}
\label{tab:lines}
\end{center}
NOTE. -- Negative equivalent widths indicate emission lines, while
positive values correspond to absorption troughs.  $\lambda_0$ is the
rest wavelength and $\lambda_{\rm obs}$ is the central wavelength at
which each feature is observed.
\end{table*}


\end{document}